\documentclass[%
reprint,
showpacs,preprintnumbers,
nofootinbib,
amsmath,amssymb,
aps,
prd,
floatfix,
]{revtex4-1}
\pdfoutput=1
\usepackage{graphicx}
\usepackage{dcolumn}
\usepackage{bm}
\newcommand{\prt}{\mu}
\newcommand{\idm}{a}
\begin{document}

\title{Predictions for Neutrinoless Double-Beta Decay in the 3+1 Sterile Neutrino Scenario}

\author{C. Giunti$^{1,2}$}
\author{E. M. Zavanin$^{1,2,3}$}
\affiliation{
$^1$
INFN, Sezione di Torino, Via P. Giuria 1, I--10125 Torino, Italy
\\
$^2$
Department of Physics, University of Torino, Via P. Giuria 1, I--10125 Torino, Italy
\\
$^3$
Instituto de F\'\i sica Gleb Wataghin, Universidade Estadual de Campinas - UNICAMP,
\\
Rua S\'ergio Buarque de Holanda, 777, 13083-859 Campinas SP Brazil
}

\date{\today}

\begin{abstract}
We present accurate predictions of the
effective Majorana mass $|m_{\beta\beta}|$
in neutrinoless double-$\beta$ decay
in the standard case of $3\nu$ mixing
and in the case of 3+1 neutrino mixing
indicated by the reactor, Gallium and LSND
anomalies.
We have taken into account the uncertainties of the
neutrino mixing parameters determined by oscillation experiments.
It is shown that the predictions for $|m_{\beta\beta}|$
in the cases of $3\nu$ and 3+1 mixing are quite different,
in agreement with previous discussions in the literature,
and that
future measurements of
neutrinoless double-$\beta$ decay
and
of the effective light neutrino mass in $\beta$ decay
or
the total mass of the three lightest neutrinos in cosmological experiments
may distinguish the $3\nu$ and 3+1 cases
if the mass ordering is determined by oscillation experiments.
We also present a relatively simple method
to determine the minimum value of
$|m_{\beta\beta}|$
in the general case of $N$-neutrino mixing.
\begin{center}
Published in: \textbf{C. Giunti, E.M. Zavanin, JHEP 1507 (2015) 171}
\end{center}
\end{abstract}

\maketitle

\section{Introduction}
\label{sec:intro}

Neutrino flavor oscillations have been observed in many
solar, reactor and accelerator experiments
(see the recent reviews in Refs.~\cite{Bellini:2013wra,Wang:2015rma}),
in agreement with the currently standard paradigm of three-neutrino ($3\nu$) mixing.
The global fits of neutrino oscillation data
in the framework of $3\nu$ mixing
\cite{Capozzi:2013csa,Forero:2014bxa,Gonzalez-Garcia:2014bfa}
give us rather precise information on the values of the
elements of the three mixing angles which
parameterize the neutrino mixing matrix
and on the values of the two independent neutrino
squared-mass differences,
the smaller ``solar'' squared-mass difference
$\Delta{m}^2_{\text{SOL}} \approx 7.5 \times 10^{-5} \, \text{eV}^2$
and the larger ``atmospheric'' squared-mass difference
$\Delta{m}^2_{\text{ATM}} \approx 2.4 \times 10^{-3} \, \text{eV}^2$.

However,
the standard $3\nu$ mixing paradigm has been challenged
by indications in favor of short-baseline oscillations
generated by a new larger squared-mass difference
$\Delta{m}^2_{\text{SBL}} \sim 1 \, \text{eV}^2$:
the reactor antineutrino anomaly
\cite{Mention:2011rk},
which is a deficit of the rate of $\bar\nu_{e}$ observed in several
short-baseline reactor neutrino experiments
in comparison with that expected from the latest calculation of
the reactor neutrino fluxes
\cite{Mueller:2011nm,Huber:2011wv};
the Gallium neutrino anomaly
\cite{Abdurashitov:2005tb,Laveder:2007zz,Giunti:2006bj,Giunti:2010zu,Giunti:2012tn},
consisting in a short-baseline disappearance of $\nu_{e}$
measured in the
Gallium radioactive source experiments
GALLEX
\cite{Kaether:2010ag}
and
SAGE
\cite{Abdurashitov:2009tn};
the signal of short-baseline
$\bar\nu_{\mu}\to\bar\nu_{e}$
oscillations observed in the LSND experiment
\cite{Athanassopoulos:1995iw,Aguilar:2001ty}.
The simplest extension of $3\nu$ mixing which can describe
these short-baseline oscillations
taking into account other constraints is the
3+1 mixing scheme
\cite{Kopp:2013vaa,Giunti:2013aea},
in which there is an additional massive neutrino at the eV scale
and
the masses of the three standard neutrinos are much smaller.
Since from the LEP measurement of the invisible width of the $Z$ boson
we know that there are only three active neutrinos
(see Ref.~\cite{Agashe:2014kda}),
in the flavor basis the additional massive neutrino corresponds to
a sterile neutrino
\cite{Pontecorvo:1968fh},
which does not have standard weak interactions.

A fundamental questions that remains open is:
are neutrinos Dirac or Majorana particles?
This question cannot be investigated in neutrino oscillation experiments,
where the total lepton number is conserved and there is no difference
between Dirac neutrinos with a conserved total lepton number
and truly neutral Majorana neutrinos,
which do not have a conserved total lepton number.
The most promising process which can reveal the Majorana nature of neutrinos
is neutrinoless double-beta decay,
in which the total lepton number changes by two units
(see the recent review in Ref.~\cite{Bilenky:2014uka}).

From the present knowledge of the neutrino squared-mass differences and mixing angles
it is possible to predict the possible range of values for the
effective Majorana mass
$|m_{\beta\beta}|$
in neutrinoless double-beta decay
as a function of the absolute scale of neutrino masses
(see Ref.~\cite{Bilenky:2014uka}),
which is still unknown,
up to the upper bound of about 2 eV at 95\% C.L.
established by the
Mainz \cite{Kraus:2004zw} and Troitsk \cite{Aseev:2011dq}
Tritium $\beta$-decay experiments.

The introduction of a sterile neutrino at the eV mass scale can change dramatically
the prediction for the possible range of values for the
effective Majorana mass in neutrinoless double-beta decay
\cite{Goswami:2005ng,Goswami:2007kv,Barry:2011wb,Li:2011ss,Rodejohann:2012xd,Giunti:2012tn,Girardi:2013zra,Pascoli:2013fiz,Meroni:2014tba,Abada:2014nwa}.
In this paper we present accurate predictions for $|m_{\beta\beta}|$
taking into account the results of the global fit
of solar, atmospheric and long-baseline reactor and accelerator
neutrino oscillation data presented in
Ref.~\cite{Capozzi:2013csa}
and the results of an update \cite{Giunti-NeuTel2015,Gariazzo:2015rra}
of the global fit of short-baseline neutrino oscillation
data presented in Ref.~\cite{Giunti:2013aea}.
We are particularly interested to determine accurately
the conditions for which $|m_{\beta\beta}| \gtrsim 0.01 \, \text{eV}$,
which may be probed experimentally in the near future
(see Refs.~\cite{GomezCadenas:2011it,Giuliani:2012zu,Schwingenheuer:2012zs,Cremonesi:2013vla,Artusa:2014wnl,Gomez-Cadenas:2015twa}),
and the conditions for which there can be a cancellation
of the different mass contributions to $|m_{\beta\beta}|$,
which leads to an unfortunate uncertainty for the possibility
of ever observing neutrinoless double-beta decay
(unless it is induced by new interactions and/or the exchange of new particles;
see Refs.~\cite{Faessler:1999zg,Choi:2002bb,Ibarra:2010xw,Tello:2010am,Rodejohann:2011mu,delAguila:2012nu,Deppisch:2012nb,deGouvea:2013zba}).

The plan of this paper is as follows.
In Section~\ref{sec:3nu}
we discuss the predictions for
$|m_{\beta\beta}|$
in the standard $3\nu$ framework,
taking into account the two possible normal and inverted mass orderings.
In Section~\ref{sec:3p1}
we discuss how these predictions are modified
in the 3+1 mixing framework.
In Section~\ref{sec:conclusions}
we draw our conclusions.

\section{Three-Neutrino Mixing}
\label{sec:3nu}

In the standard three-neutrino ($3\nu$) mixing framework,
the effective Majorana mass in neutrinoless double-beta decay
is given by
\begin{equation}
|m_{\beta\beta}| = \left| \prt_{1} + \prt_{2} e^{i\alpha_2} + \prt_{3} e^{i\alpha_3} \right|
,
\label{mbb3nu}
\end{equation}
where
\begin{equation}
\prt_{k}
=
|U_{ek}|^2 m_{k}
\label{mkp}
\end{equation}
is the partial contribution of the massive Majorana neutrino $\nu_{k}$ with mass $m_{k}$.
The elements $U_{ek}$ of the mixing matrix,
which quantify the mixing of the electron neutrino with the three massive neutrinos,
can have unknown complex phases,
which generate the two complex phases
$\alpha_2$
and
$\alpha_3$
in Eq.~(\ref{mbb3nu}).
Since the values of these phases is completely unknown,
all predictions of the value of
$|m_{\beta\beta}|$
must take into account all the possible range of these phases,
from 0 to $2\pi$.

We use the results of the global fit
of solar, atmospheric and long-baseline reactor and accelerator
neutrino oscillation data presented in
Ref.~\cite{Capozzi:2013csa},
which are given in terms of the mixing angles
$\vartheta_{12}$,
$\vartheta_{13}$
that determine the absolute  values of the first row of the
mixing matrix $U$
in the standard parameterization:
\begin{align}
&
|U_{e1}| = \cos\vartheta_{13} \cos\vartheta_{12},
\label{3nuUe1}
\\
&
|U_{e2}| = \cos\vartheta_{13} \sin\vartheta_{12},
\label{3nuUe2}
\\
&
|U_{e3}| = \sin\vartheta_{13}.
\label{3nuUe3}
\end{align}
The results for the neutrino squared-mass differences are expressed in terms
of the solar and atmospheric squared mass differences,
which are defined by
\begin{align}
&
\Delta{m}_{\text{SOL}}^2 = \Delta{m}_{21}^2
,
\label{dmsol}
\\
&
\Delta{m}_{\text{ATM}}^2 = \frac{1}{2} \left| \Delta{m}_{31}^2 + \Delta{m}_{32}^2 \right|
,
\label{dmatm}
\end{align}
where $\Delta m^2_{jk} = m_{j}^2 - m_{k}^2$.
Given this assignment of the squared mass differences,
it is currently unknown if the ordering of the neutrino masses
is normal (NO),
such that $m_{1}<m_2<m_3$
or inverted (IO),
such that $m_3<m_{1}<m_2$.
We discuss these two cases separately in the following subsections.

\begin{figure}[t!]
\centering
\includegraphics*[width=\linewidth]{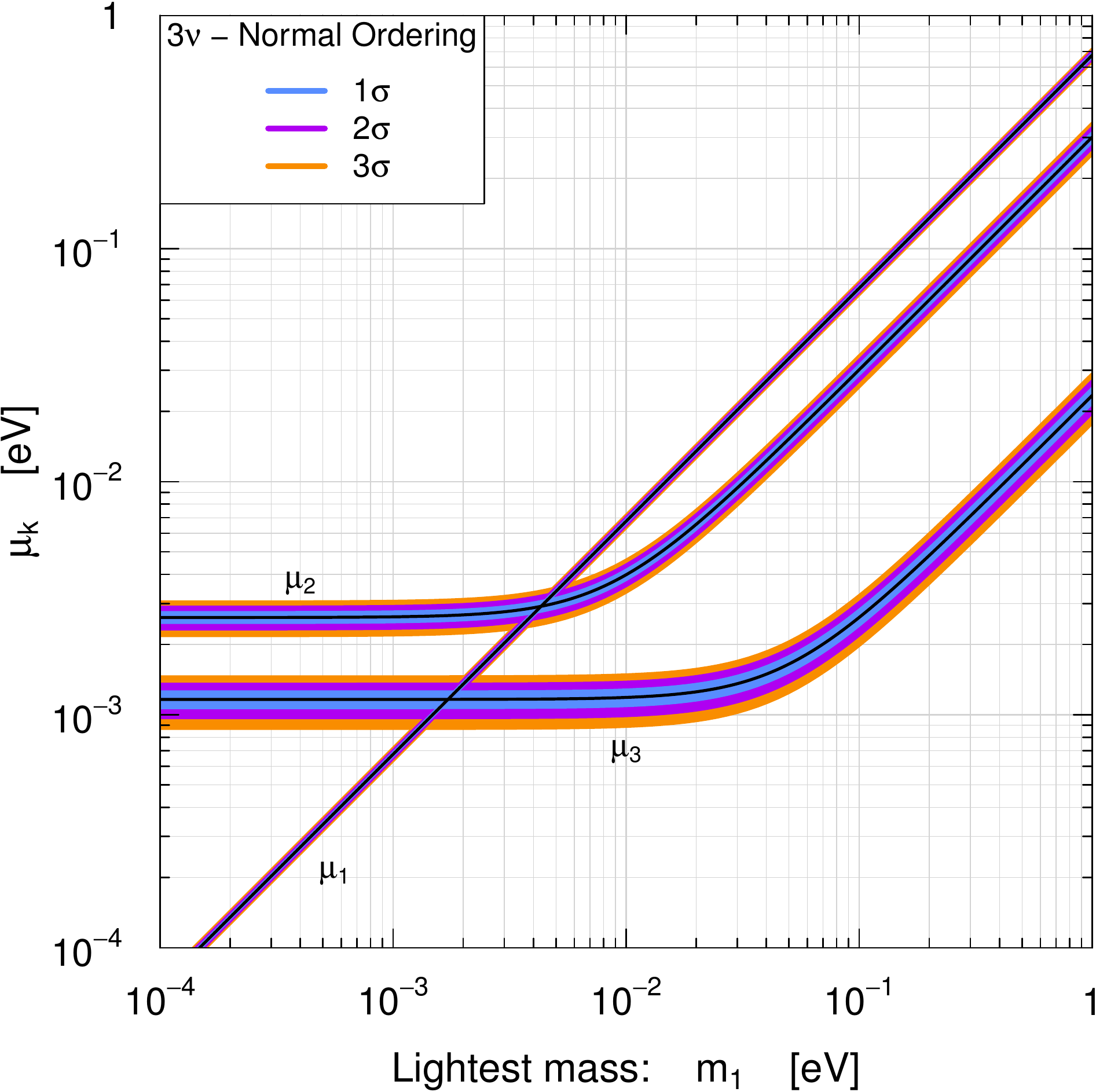}
\caption{Best-fit values (b.f.)
and
$1\sigma$,
$2\sigma$ and
$3\sigma$ allowed intervals
of the three partial mass contributions to $|m_{\beta\beta}|$ in Eq.~(\ref{mbb3nu})
as functions of
the lightest mass $m_{1}$
in the case of $3\nu$ mixing with Normal Ordering.}
\label{3nuNOpartial}
\end{figure}

\begin{figure}[t!]
\centering
\includegraphics*[width=\linewidth]{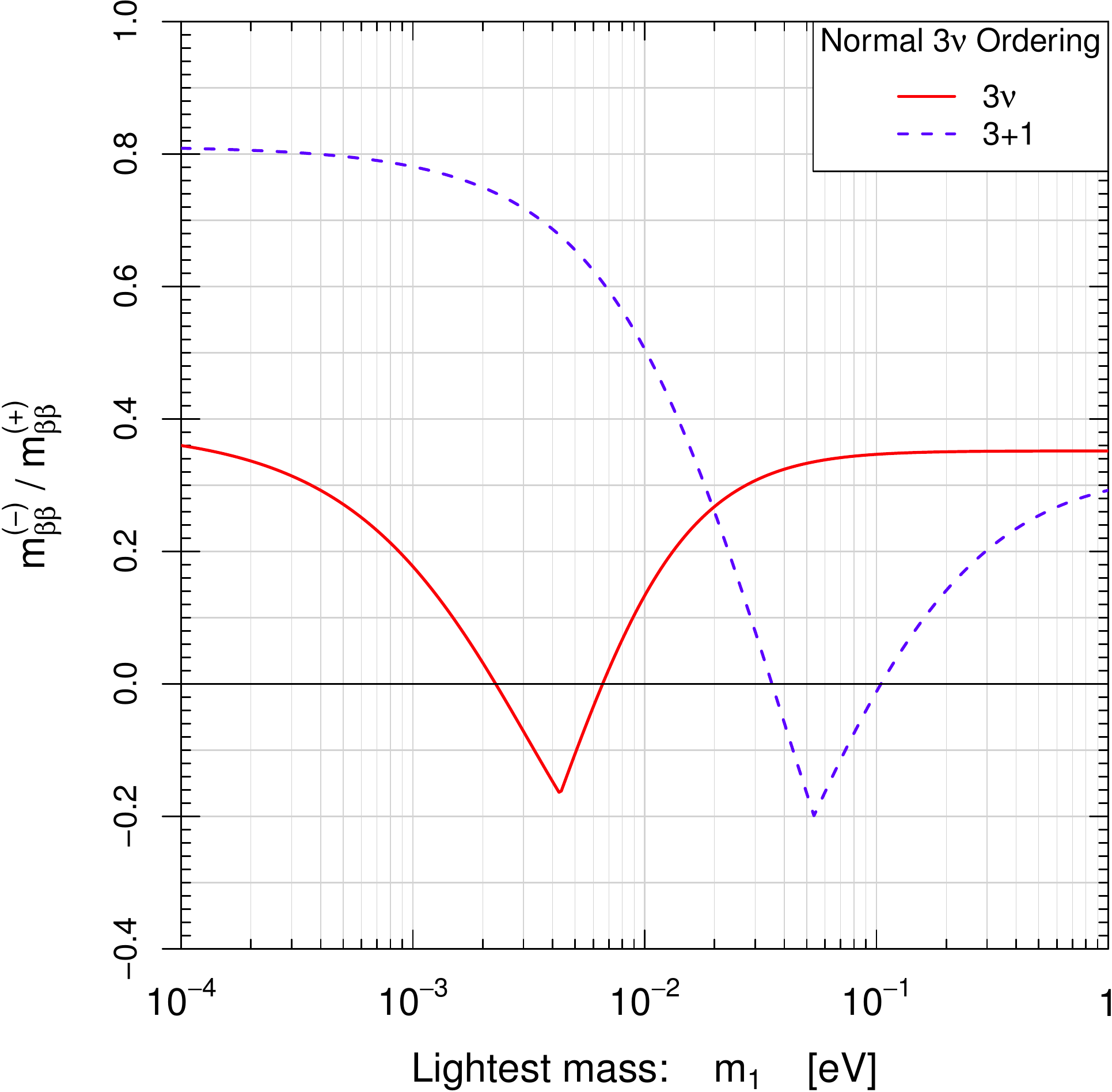}
\caption{
Ratio
$m_{\beta\beta}^{(-)}/m_{\beta\beta}^{(+)}$
(see Eq.~(\ref{mbbpm}))
as a function of $m_{1}$ for the best-fit values of the
partial mass contributions
in the case of $3\nu$ and 3+1 mixing with Normal Ordering
of the three lightest neutrinos.
}
\label{ratNOmbbvsmin}
\end{figure}

\begin{figure}[t!]
\centering
\includegraphics*[width=\linewidth]{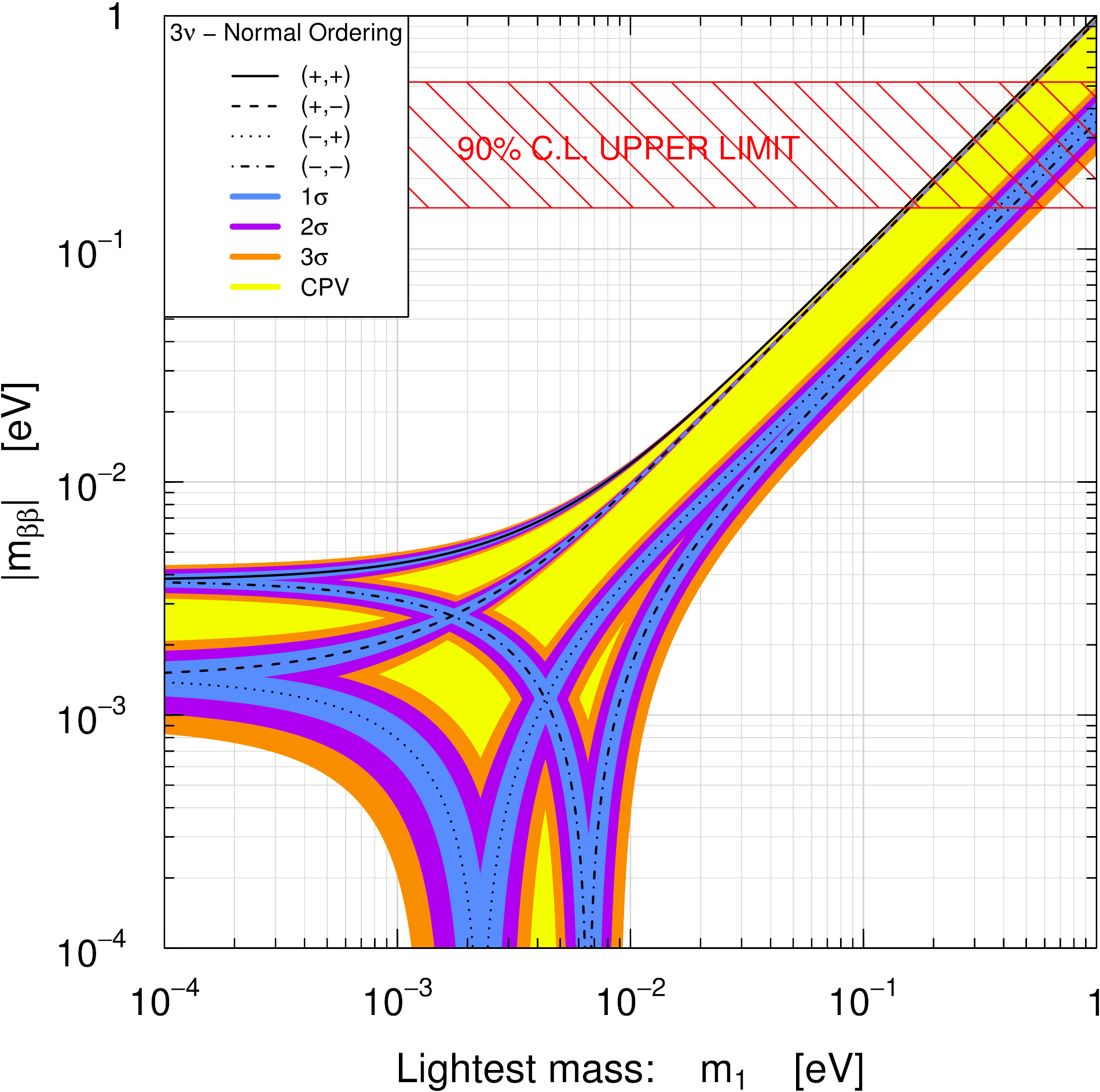}
\caption{
Value of the effective Majorana mass $|m_{\beta\beta}|$
as a function of the lightest neutrino mass $m_{1}$
in the case of $3\nu$ mixing with Normal Ordering.
The signs in the legend indicate the signs of
$e^{i\alpha_{2}}, e^{i\alpha_{3}} = \pm1$
for the four possible cases in which CP is conserved.
The intermediate yellow region is allowed only in the case of CP violation.
The 90\% upper limit is explained in the main text.
}
\label{3nuNOmbbvsmin}
\end{figure}

\begin{figure}[t!]
\centering
\includegraphics*[width=\linewidth]{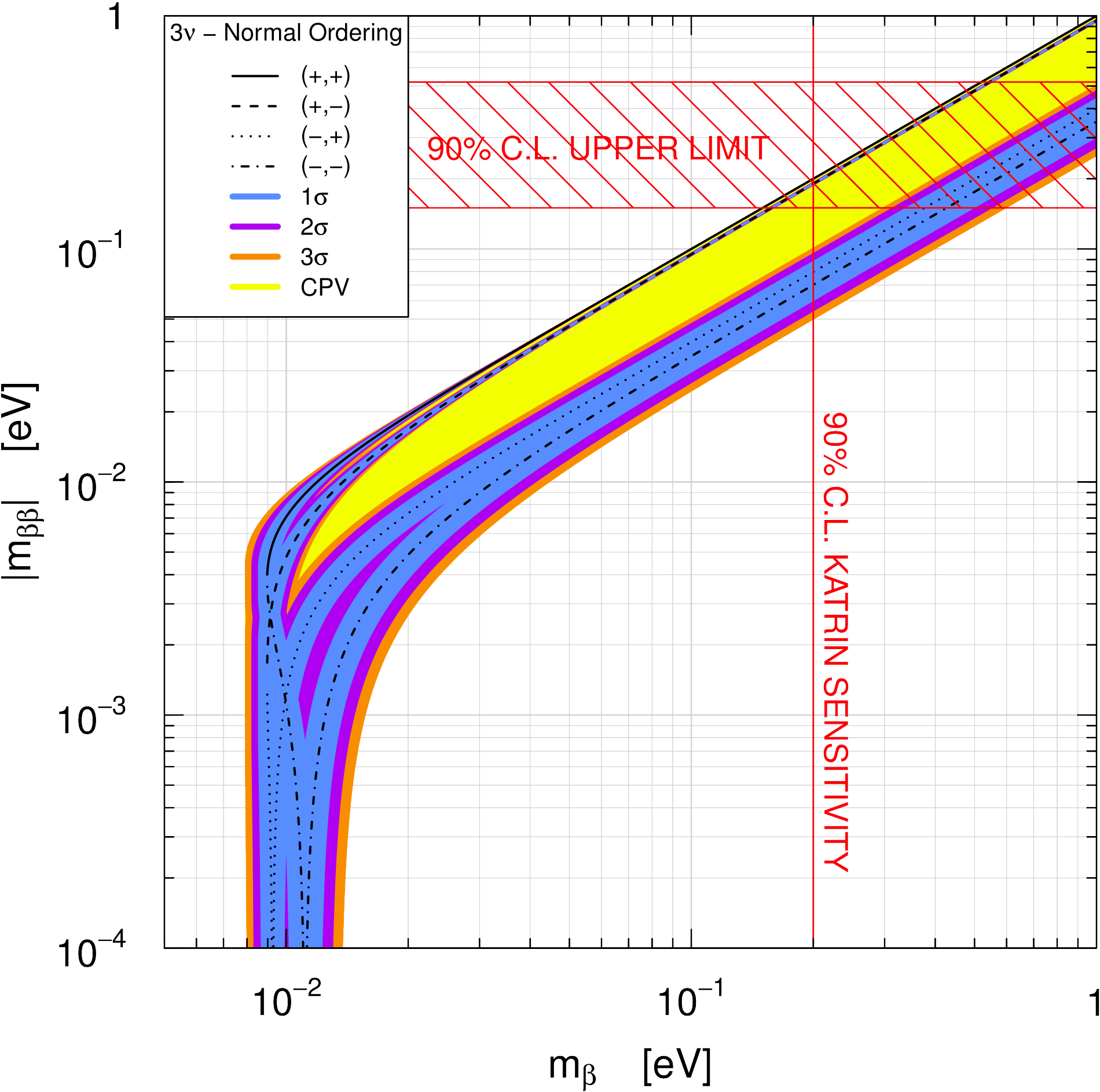}
\caption{
Value of the effective Majorana mass $|m_{\beta\beta}|$
as a function of effective electron neutrino mass $m_{\beta}$ in Eq.~(\ref{mb})
in the case of $3\nu$ mixing with Normal Ordering.
The legend is explained in the caption of Fig.~\ref{3nuNOmbbvsmin}.
The limits are explained in the main text.
}
\label{3nuNOmbbvsmb}
\end{figure}

\begin{figure}[t!]
\centering
\includegraphics*[width=\linewidth]{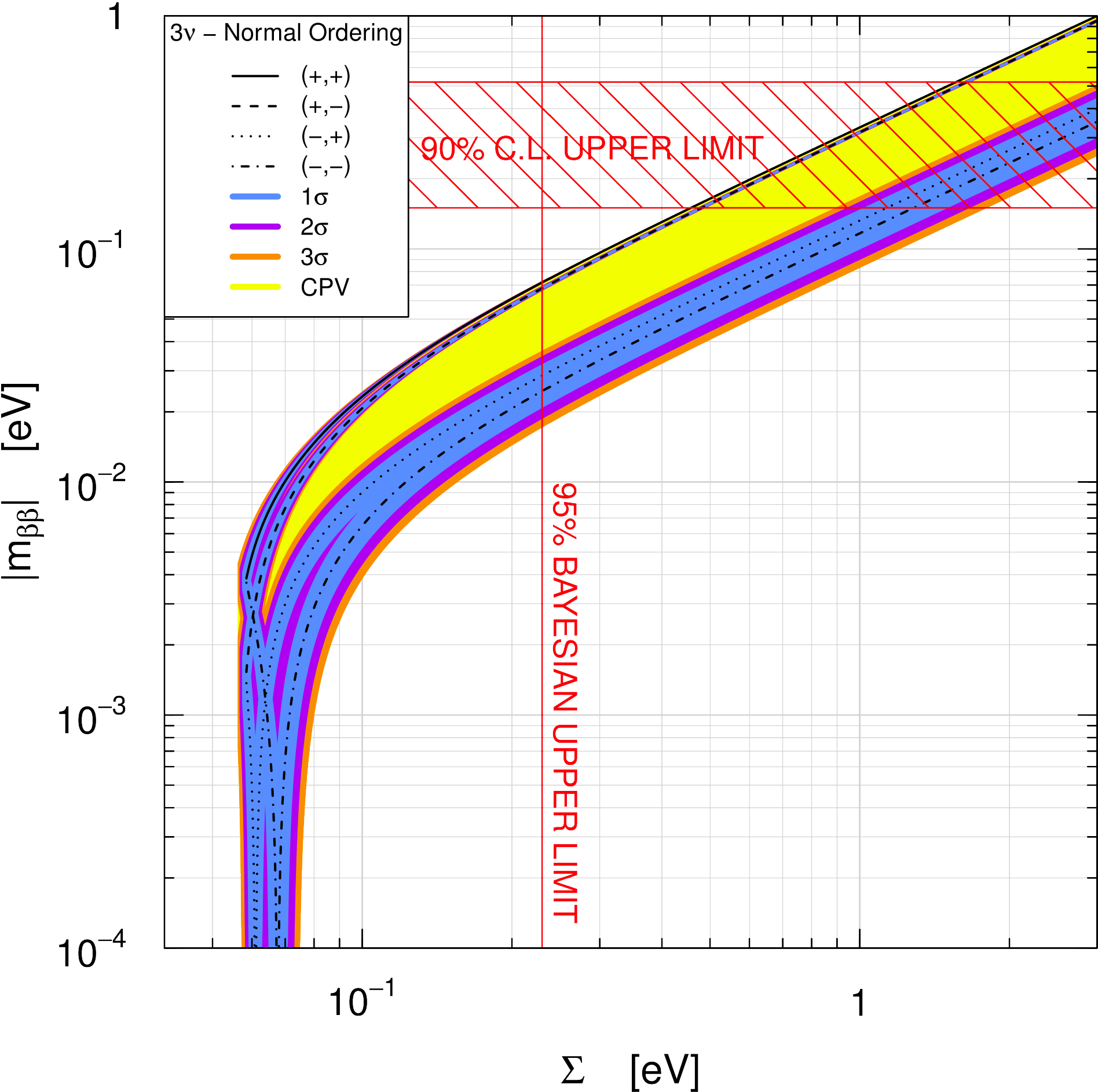}
\caption{
Value of the effective Majorana mass $|m_{\beta\beta}|$
as a function of sum of the neutrino masses $\Sigma$ in Eq.~(\ref{sum})
in the case of $3\nu$ mixing with Normal Ordering.
The legend is explained in the caption of Fig.~\ref{3nuNOmbbvsmin}.
The limits are explained in the main text.
}
\label{3nuNOmbbvssum}
\end{figure}

\begin{table}[b!]
\caption{Ranges of
$m_{1}$,
$m_{\beta}$ and
$\Sigma$
for which there can be a complete cancellation of the
three partial mass contributions to $|m_{\beta\beta}|$
for the best-fit values (b.f.) of the oscillation parameters
and at
$1\sigma$,
$2\sigma$ and
$3\sigma$
in the case of $3\nu$ mixing with Normal Ordering.
}
\centering
\renewcommand{\arraystretch}{1.2}
\begin{tabular}{ccccc}
\hline
\hline
&
b.f.
&
$1\sigma$
&
$2\sigma$
&
$3\sigma$
\\
\hline
$m_{1}\,[10^{-3}\,\text{eV}]$
&
$2.3 - 6.6$
&
$1.9 - 7.2$
&
$1.6 - 8.0$
&
$1.3 - 9.0$
\\
$m_{\beta}\,[10^{-2}\,\text{eV}]$
&
$0.9 - 1.1$
&
$0.9 - 1.2$
&
$0.8 - 1.3$
&
$0.8 - 1.4$
\\
$\Sigma\,[10^{-2}\,\text{eV}]$
&
$6.1 - 6.8$
&
$5.9 - 7.0$
&
$5.8 - 7.2$
&
$5.7 - 7.4$
\\
\hline
\hline
\end{tabular}
\label{tab:3nu-nor}
\end{table}

\subsection{Normal Ordering}
\label{sub:3nuNO}

In order to study the case of
Normal Ordering (NO),
we express the neutrino masses in terms of the lightest neutrino mass
$m_{\text{min}}$:
\begin{align}
&
m_{1} = m_{\text{min}}
,
\label{m1NO3nu}
\\
&
m_{2} = \sqrt{m_{\text{min}}^2 + \Delta{m}_{\text{SOL}}^2}
,
\label{m2NO3nu}
\\
&
m_{3} = \sqrt{m_{\text{min}}^2 + \Delta{m}_{\text{ATM}}^2 + \Delta{m}_{\text{SOL}}^2/2}
.
\label{m3NO3nu}
\end{align}

Figure~\ref{3nuNOpartial}
shows the best-fit values
and the
$1\sigma$,
$2\sigma$ and
$3\sigma$ allowed intervals
of the three partial mass contributions to $|m_{\beta\beta}|$ in Eq.~(\ref{mbb3nu})
as functions of
the lightest mass $m_{1}$.
We calculated the confidence intervals using the $\chi^2$ function
\begin{align}
\chi^2_{3\nu}
=
\null & \null
\chi^2(\Delta{m}_{\text{SOL}}^2)
+
\chi^2(\Delta{m}_{\text{ATM}}^2)
\nonumber
\\
\null & \null
+
\chi^2(\sin^2\vartheta_{12})
+
\chi^2(\sin^2\vartheta_{13})
,
\label{chi3nu}
\end{align}
with the partial $\chi^2$'s extracted from Fig.~3 of Ref.~\cite{Capozzi:2013csa},
neglecting possible small correlations of the four mixing parameters\footnote{
The only significant correlations
discussed in Ref.~\cite{Capozzi:2013csa}
are those which involve the mixing angle
$\vartheta_{23}$,
which is irrelevant for neutrinoless double-$\beta$ decay,
and the Dirac phase, whose effect in neutrinoless double-$\beta$ decay
is masked by the two unknown Majorana phases.
}.
For each value of $m_{1}$
we calculated the confidence intervals for one degree of freedom.
Although this method is in principle better than the method based on the
propagation of errors of the parameters used by many authors,
in practice it leads to similar results,
because the $\chi^2$'s of
$\Delta{m}_{\text{SOL}}^2$,
$\Delta{m}_{\text{ATM}}^2$,
$\sin^2\vartheta_{12}$ and
$\sin^2\vartheta_{13}$
in Fig.~3 of Ref.~\cite{Capozzi:2013csa}
are very well approximated by quadratic functions,
which correspond to Gaussian uncertainties for which the
method of propagation of errors is valid.

From Fig.~\ref{3nuNOpartial}
one can see that for
$m_{1} \lesssim 2 \times 10^{-3} \, \text{eV}$
the contribution of $\prt_{2}$ is dominant
and cannot be canceled by the smaller contributions of
$\prt_{1}$
and
$\prt_{3}$
for any values of the relative phase differences.
In the interval
$ m_{1} \approx (2-7) \times 10^{-3} \, \text{eV} $
cancellations are possible,
mainly between
$\prt_{1}$
and
$\prt_{2}$
which have similar values,
with the smaller contribution of $\prt_{3}$,
which is about 2.3 times smaller than $\prt_{2}$.
For $ m_{1} \gtrsim 7 \times 10^{-3} \, \text{eV} $
again there cannot be a complete cancellation,
because the contribution of
$\prt_{1}$ is dominant.

The result for $|m_{\beta\beta}|$
is shown in Fig.~\ref{3nuNOmbbvsmin},
where we have plotted separately the
allowed bands for the four possible cases in which CP is conserved
($\alpha_{2}, \alpha_{3} = 0, \pi$)
and the coefficients of the contributions are real.
These are the extreme cases which determine the minimum and maximum values
of $|m_{\beta\beta}|$.
The areas between the CP-conserving curves
correspond to values of
$|m_{\beta\beta}|$ which are allowed only in the case of CP violation
\cite{Bilenky:2001rz,Pascoli:2002qm,Joniec:2004mx,Pascoli:2005zb,Simkovic:2012hq,Pascoli:2013fiz}
(although there is no manifest CP violation \cite{deGouvea:2002gf}).

Figure~\ref{3nuNOmbbvsmin} shows also the
90\% C.L. upper limit band for $|m_{\beta\beta}|$
estimated in Ref.~\cite{Bilenky:2014uka}
from the results of the
KamLAND-Zen experiment \cite{Gando:2012zm}
taking into account the uncertainties of the nuclear matrix element calculations.
The reliability of this upper limit is supported by the upper limits with the same order of magnitude
following from the results of the
Heidelberg-Moscow \cite{Klapdor-Kleingrothaus:2001yx},
IGEX \cite{Aalseth:2002rf},
GERDA \cite{Agostini:2013mzu},
NEMO-3 \cite{Arnold:2013dha},
CUORICINO \cite{Andreotti:2010vj}, and
EXO \cite{Albert:2014awa}
experiments.

From Fig.~\ref{3nuNOmbbvsmin} one can see that,
in agreement with the discussion above,
there can be a complete cancellation of the
three partial mass contributions to $|m_{\beta\beta}|$
for $m_{1}$ in the intervals given in Tab.~\ref{tab:3nu-nor}
at different confidence levels.

The exact determination of the region in which there
can be a complete cancellation of the partial mass contributions to $|m_{\beta\beta}|$
in the general case of $N$-neutrino mixing
can be done in the following relatively simple way\footnote{
Other ways are discussed in Refs.~\cite{Vissani:1999tu,Girardi:2013zra,Xing:2014yka}.
}.
For each value of the lightest mass $m_1$
let us denote by $\idm$
the index of the largest mass contribution,
i.e.
\begin{equation}
\prt_{\idm} \geq \prt_{k}
\quad
\text{for}
\quad
k \neq \idm
.
\label{prtmax}
\end{equation}
Then, we can consider the quantities
\begin{equation}
m_{\beta\beta}^{(\pm)}
=
\prt_{\idm}
\pm
\sum_{k \neq \idm}
\prt_{k}
.
\label{mbbpm}
\end{equation}
The quantity
$m_{\beta\beta}^{(+)}$
is always positive and
represents the most favorable case,
in which all the mass contribution add with the same phase,
giving the maximum value of
$|m_{\beta\beta}|$
for any value of the unknown phases:
\begin{equation}
|m_{\beta\beta}|_{\text{max}}
=
m_{\beta\beta}^{(+)}
.
\label{mbbmax}
\end{equation}
The quantity
$m_{\beta\beta}^{(-)}$
represents the extreme case in which the phases of all the other partial mass contributions
are equal and opposite to the phase of the largest mass contribution.
It is evident that if
$m_{\beta\beta}^{(-)} > 0$,
the value of $m_{\beta\beta}$ gives the minimum possible value of
$|m_{\beta\beta}|$
for any value of the unknown phases,
because it corresponds to the maximal cancellation between
$\prt_{\idm}$
and
the maximum
$\sum_{k \neq \idm} \prt_{k}$
of the other partial mass contributions.
On the other hand,
if
\begin{equation}
m_{\beta\beta}^{(-)} \leq 0
,
\label{mbb5}
\end{equation}
there is an intermediate value of the phases which gives
$|m_{\beta\beta}|=0$.
This can be seen clearly by writing
$|m_{\beta\beta}|$
as
\begin{equation}
|m_{\beta\beta}|
=
\left|
\prt_{\idm}
+
e^{i\alpha'}
\prt'
\right|
,
\label{mbb1}
\end{equation}
with
\begin{equation}
\prt'
=
\left|
\sum_{k \neq \idm}
e^{i\xi_{k}}
\prt_{k}
\right|
,
\label{mbb2}
\end{equation}
with an unknown phase $\alpha'$ and
$N-2$ unknown phases
$\xi_{k}$,
of which one can be fixed to zero.
The only possibility to have
$|m_{\beta\beta}|=0$
can be realized for $\alpha'=\pi$
if $\prt'=\prt_{\idm}$.
This equality can occur only if
\begin{equation}
\prt'_{\text{min}}
\leq
\prt_{\idm}
\leq
\prt'_{\text{max}}
,
\label{mbb3}
\end{equation}
where
$\prt'_{\text{min}}$
and
$\prt'_{\text{max}}$
are the minimum and maximum values of
$\prt'$
for any value of the unknown phases
$\xi_{k}$.
Since we have always $\prt'_{\text{min}}\leq\prt_{\idm}$ because of Eq.~(\ref{prtmax})
and
\begin{equation}
\prt'_{\text{max}}
=
\sum_{k \neq \idm}
\prt_{k}
,
\label{mbb4}
\end{equation}
the inequality in Eq.~(\ref{mbb5})
is the necessary and sufficient condition for having
$|m_{\beta\beta}|=0$
for some value of the unknown phases.
Hence,
the minimum value of $|m_{\beta\beta}|$
is given by
\begin{equation}
|m_{\beta\beta}|_{\text{min}}
=
\max\!\left[
m_{\beta\beta}^{(-)}, 0
\right]
,
\label{mbbmin}
\end{equation}
which can also be written as
\cite{Vissani:1999tu}\footnote{
Equation~(\ref{mbbminv})
is the generalization to $N$-neutrino mixing
of that obtained with a different proof in Ref.~\cite{Vissani:1999tu}
in the case of $3\nu$ mixing.
}
\begin{equation}
|m_{\beta\beta}|_{\text{min}}
=
\max\!\left[
2 \prt_{k}
-
|m_{\beta\beta}|_{\text{max}}
,
0
\right]
.
\label{mbbminv}
\end{equation}

Fig.~\ref{ratNOmbbvsmin}
shows the value of the ratio
$m_{\beta\beta}^{(-)}/m_{\beta\beta}^{(+)}$
as a function of $m_{1}$ for the best-fit values of the
partial mass contributions,
which is sufficient for the determination of the interval of
$m_{1}$ for which there can be a complete cancellation
of the partial mass contributions.
One can see that in the case of $3\nu$ mixing
$m_{\beta\beta}^{(-)}$
is negative
and it is possible that $|m_{\beta\beta}| = 0$
only in the interval of $m_{1}$
given in Tab.~\ref{tab:3nu-nor}.

Let us now consider the opposite possibility that $|m_{\beta\beta}|$
is larger than about 0.01 eV,
which is a value that
may be explored experimentally in the near future.
From Fig.~\ref{3nuNOmbbvsmin}
one can see that
$|m_{\beta\beta}| \gtrsim 0.01 \, \text{eV}$
can be realized only for $m_{1} \gtrsim 0.008 \, \text{eV}$.
This range of $m_{1}$ corresponds to almost degenerate
$m_{1}$ and $m_{2}$,
because
$\sqrt{\Delta{m}^2_{\text{SOL}}} \approx 8.7 \times 10^{-3} \, \text{eV}$.
Hence, it will be very difficult to
measure $|m_{\beta\beta}|$
if there is a normal hierarchy of neutrino masses
($m_{1} \ll m_{2} \ll m_{3}$)
for any value of the unknown phases
$\alpha_2$
and
$\alpha_3$
in Eq.~(\ref{mbb3nu}).

One can also see from Fig.~\ref{3nuNOmbbvsmin} that
$|m_{\beta\beta}| \gtrsim 0.01 \,\text{eV}$
is realized independently of
the values of the unknown phases
$\alpha_2$
and
$\alpha_3$
for
$m_{1} \gtrsim 0.04 \, \text{eV}$,
which is close to the region
$m_{1} \gtrsim \sqrt{\Delta{m}^2_{\text{ATM}}} \approx 0.05 \, \text{eV}$
in which all the three neutrino masses are quasidegenerate.

Figure~\ref{3nuNOmbbvsmin} gives a clear view of the possible values
of $|m_{\beta\beta}|$ depending on the scale of the lightest mass $m_{1}$,
but it is of little practical usefulness,
because it will be very difficult to measure directly the value of $m_{1}$.
In practice,
the investigation of the absolute values of neutrino masses
is performed through the measurements
of the effective electron neutrino mass
\begin{equation}
m_{\beta} = \sqrt{|U_{e1}|^2 m_{1}^2 + |U_{e2}|^2 m_2^2 + |U_{e3}|^2 m_3^2}
\label{mb}
\end{equation}
in $\beta$-decay experiments \cite{Kraus:2004zw,Aseev:2011dq}
and through the measurement of the sum of the neutrino masses
\begin{equation}
\Sigma = m_{1} + m_2 + m_3
\label{sum}
\end{equation}
in cosmological experiments
(see, for example, Ref.~\cite{Ade:2015xua}).
Hence,
it is useful to calculate the allowed regions
in the
$m_{\beta}$--$|m_{\beta\beta}|$
and
$\Sigma$--$|m_{\beta\beta}|$
planes
\cite{Barger:1999na,Matsuda:2000iw,Barger:2002xm,Fogli:2004as},
which are shown in Figs.~\ref{3nuNOmbbvsmb} and \ref{3nuNOmbbvssum}.
In this case,
the confidence intervals are calculated using the $\chi^2$ function in Eq.~(\ref{chi3nu})
with two degrees of freedom.
We have plotted separately the
allowed bands for the four possible cases in which CP is conserved
($\alpha_{2}, \alpha_{3} = 0, \pi$),
in order to show the regions
in which CP is violated.
Potentially the possibility of measuring values of $|m_{\beta\beta}|$
and
$m_{\beta}$ or $\Sigma$
in these regions is very exciting for the discovery of CP violation
generated by the Majorana phases\footnote{
The phases
$\alpha_2$
and
$\alpha_3$
in Eq.~(\ref{mbb3nu})
depend on the values of one Dirac phase and two Majorana phases in the
neutrino mixing matrix
(see Ref.~\cite{Giunti:2007ry}).
The Dirac phase can be measured in neutrino oscillation experiments
and there is some indication on its value
\cite{Capozzi:2013csa,Forero:2014bxa,Gonzalez-Garcia:2014bfa}.
On the other hand,
the values of the two Majorana phases
can be measured only in lepton-number violating processes
such as neutrinoless double-$\beta$ decay.
In the future,
if the Dirac phase will be measured,
CP violation in neutrinoless double-$\beta$ decay
may provide information on the Majorana phases.
},
but in practice such measurement is very difficult because it would require
a precision which seems to be beyond what can be currently envisioned
\cite{Barger:2002vy,Nunokawa:2002iv,Minakata:2014jba,Dell'Oro:2014yca},
especially taking into account the current uncertainty of the
calculation of the nuclear matrix element in
neutrinoless double-$\beta$ decay
(see Refs.~\cite{Vergados:2012xy,Suhonen:2013zda,Yoshida:2013jh,Bilenky:2014uka}).

Figures~\ref{3nuNOmbbvsmb} and \ref{3nuNOmbbvssum} show the
same 90\% C.L. upper limit band for $|m_{\beta\beta}|$
as in Fig.~\ref{3nuNOmbbvsmin}.
In addition,
Fig.~\ref{3nuNOmbbvsmb} shows\footnote{
The most stringent current upper limits on $m_{\beta}$ obtained in the
Mainz ($m_{\beta} < 2.3 \, \text{eV}$ at 95\% C.L.) \cite{Kraus:2004zw}
and
Troitsk ($m_{\beta} < 2.1 \, \text{eV}$ 95\% C.L.) \cite{Aseev:2011dq}
experiments
are out of the scale in Fig.~\ref{3nuNOmbbvsmb}.
}
the 90\% C.L. sensitivity on
$m_{\beta}$
of the KATRIN experiment \cite{Mertens:2015ila},
which is scheduled to start data taking in 2016,
and
Fig.~\ref{3nuNOmbbvssum} shows the 95\% bayesian upper limit on
$\Sigma$
obtained by the Planck collaboration
\cite{Ade:2015xua}.

The intervals of
$m_{\beta}$ and $\Sigma$
for which there can be a complete cancellation of the
three partial mass contributions to $|m_{\beta\beta}|$
are given in Tab.~\ref{tab:3nu-nor}.
On the other hand,
from  Figs.~\ref{3nuNOmbbvsmb} and \ref{3nuNOmbbvssum}
one can see that
$|m_{\beta\beta}| \gtrsim 0.01 \,\text{eV}$
for any value of the unknown phases
$\alpha_2$
and
$\alpha_3$
for
$m_{\beta} \gtrsim 0.05 \, \text{eV}$
and
$\Sigma \gtrsim 0.15 \, \text{eV}$.
The $3\sigma$ lower bounds for
$m_{\beta}$ and $\Sigma$
are,
respectively,
$0.8 \times 10^{-2} \, \text{eV}$
and
$5.6 \times 10^{-2} \, \text{eV}$.

\begin{figure}[t!]
\centering
\includegraphics*[width=\linewidth]{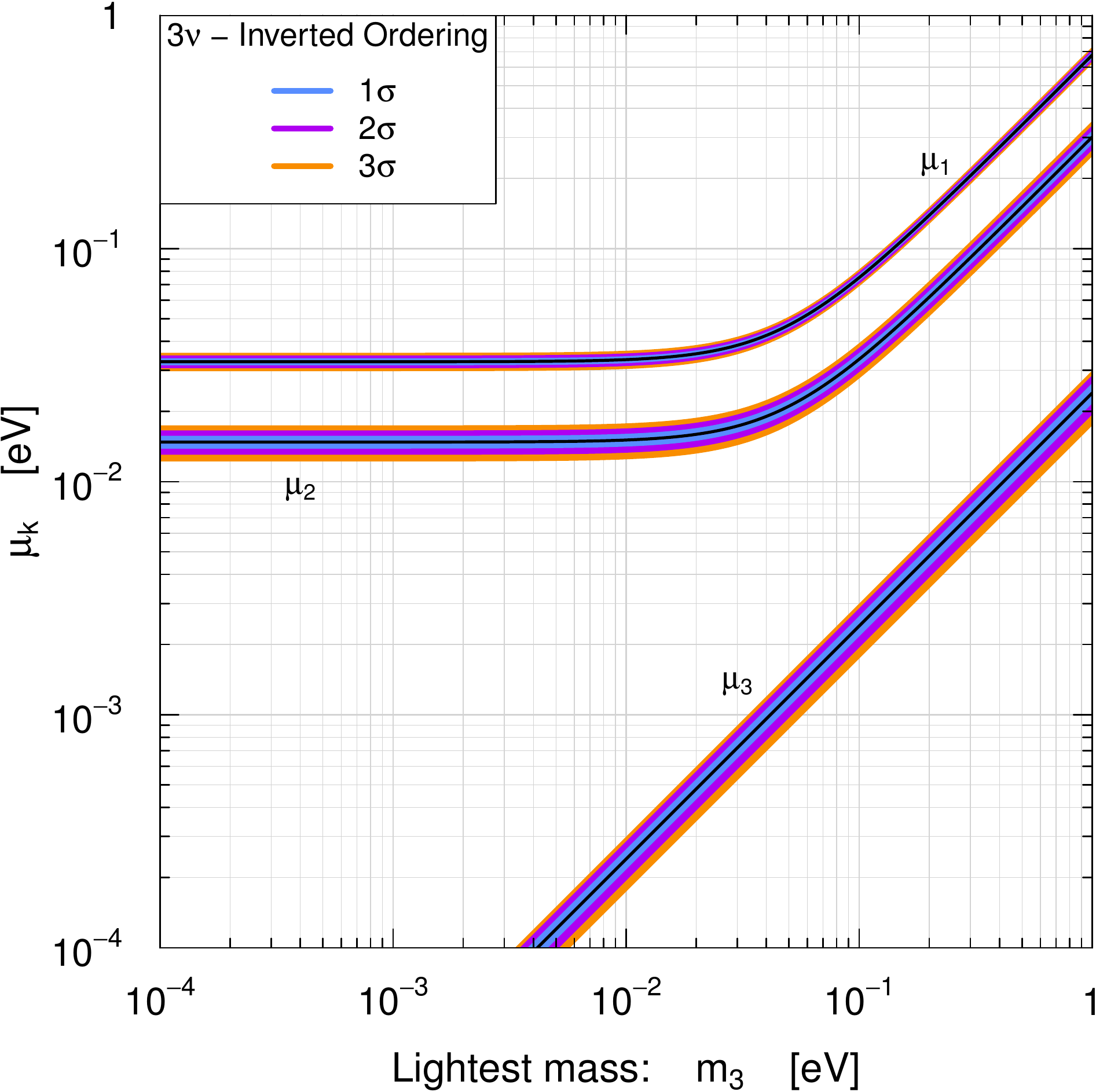}
\caption{
Best-fit values (b.f.)
and
$1\sigma$,
$2\sigma$ and
$3\sigma$ allowed intervals
of the three partial mass contributions to $|m_{\beta\beta}|$ in Eq.~(\ref{mbb3nu})
as functions of
the lightest mass $m_{3}$
in the case of $3\nu$ mixing with Inverted Ordering.}
\label{3nuIOpartial}
\end{figure}

\begin{figure}[t!]
\centering
\includegraphics*[width=\linewidth]{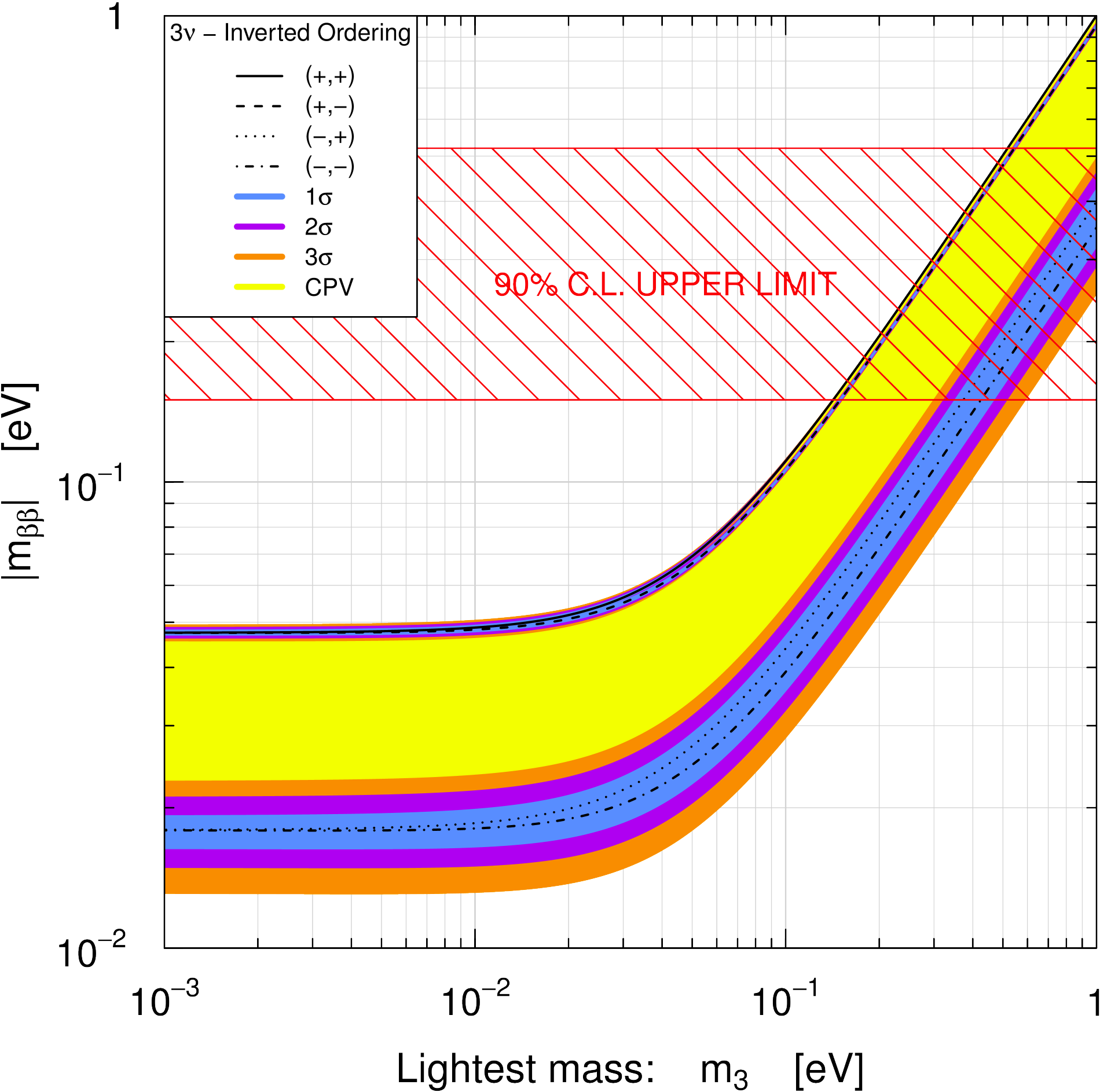}
\caption{
Value of the effective Majorana mass $|m_{\beta\beta}|$ as a function of lightest neutrino mass in the three neutrino case for the Inverted Ordering.
The legend is explained in the caption of Fig.~\ref{3nuNOmbbvsmin}.
The 90\% upper limit is explained in Section~\ref{sub:3nuNO}.
}
\label{3nuIOmbbvsmin}
\end{figure}

\begin{figure}[t!]
\centering
\includegraphics*[width=\linewidth]{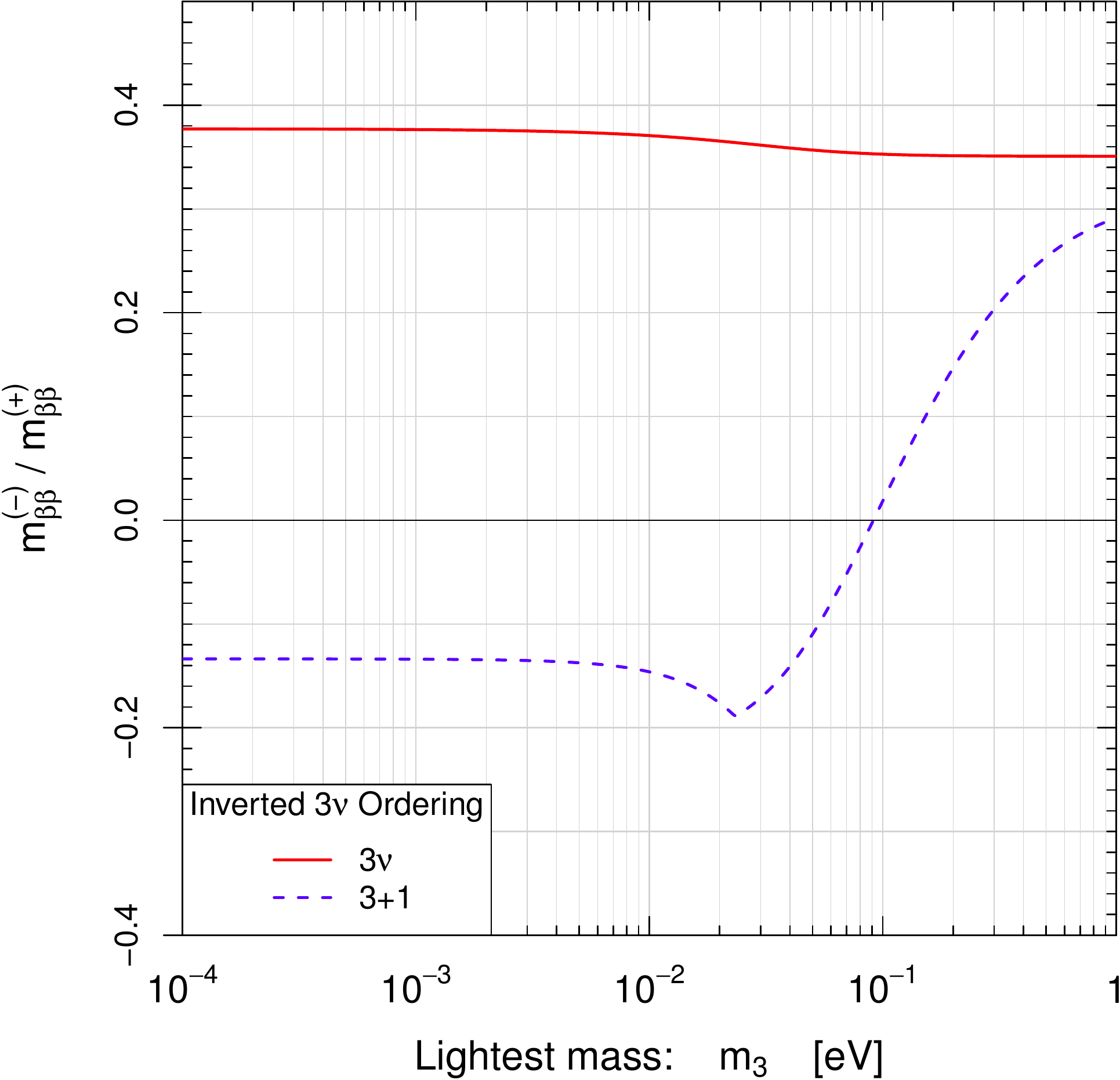}
\caption{
Ratio
$m_{\beta\beta}^{(-)}/m_{\beta\beta}^{(+)}$
(see Eq.~(\ref{mbbpm}))
as a function of $m_{3}$ for the best-fit values of the
partial mass contributions
in the case of $3\nu$ and 3+1 mixing with Inverted Ordering
of the three lightest neutrinos.
}
\label{ratIOmbbvsmin}
\end{figure}

\begin{figure}[t!]
\centering
\includegraphics*[width=\linewidth]{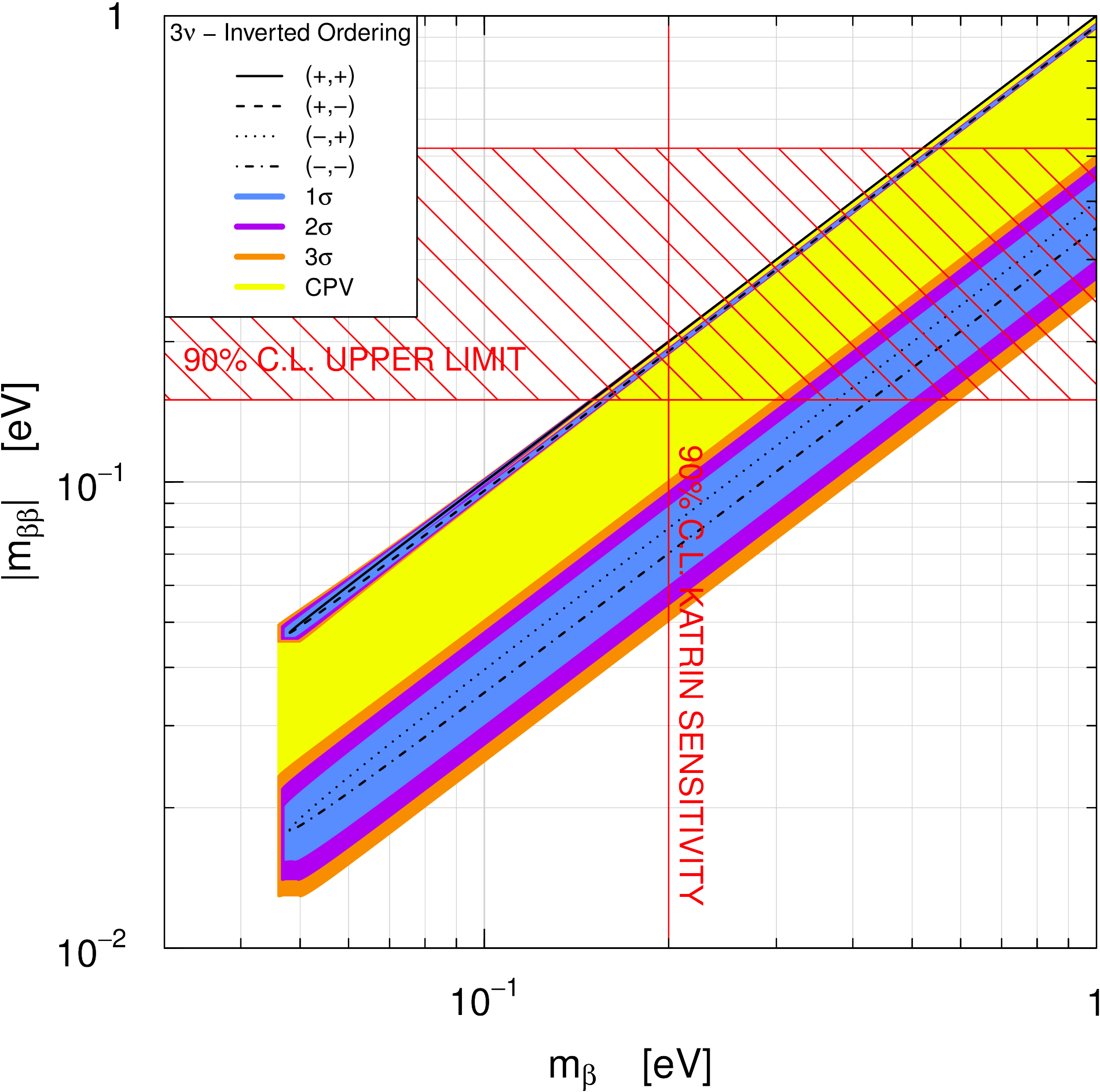}
\caption{
Value of the effective Majorana mass $|m_{\beta\beta}|$
as a function of effective electron neutrino mass $m_{\beta}$ in Eq.~(\ref{mb})
in the case of $3\nu$ mixing with Inverted Ordering.
The legend is explained in the caption of Fig.~\ref{3nuNOmbbvsmin}.
The limits are explained in in Section~\ref{sub:3nuNO}.
}
\label{3nuIOmbbvsmb}
\end{figure}

\begin{figure}[t!]
\centering
\includegraphics*[width=\linewidth]{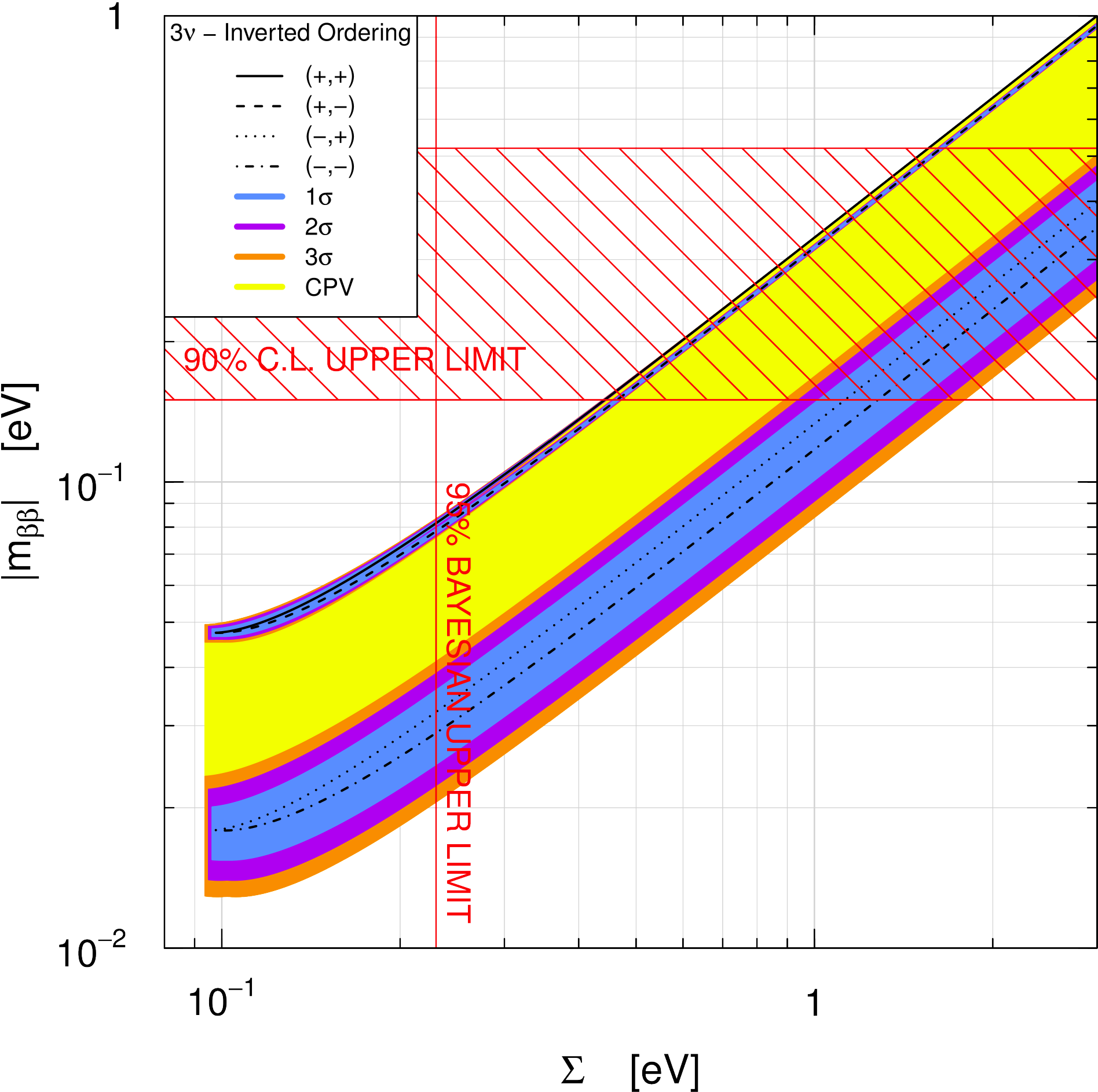}
\caption{
Value of the effective Majorana mass $|m_{\beta\beta}|$
as a function of sum of the neutrino masses $\Sigma$ in Eq.~(\ref{sum})
in the case of $3\nu$ mixing with Inverted Ordering.
The legend is explained in the caption of Fig.~\ref{3nuNOmbbvsmin}.
The limits are explained in in Section~\ref{sub:3nuNO}.
}
\label{3nuIOmbbvssum}
\end{figure}

\subsection{Inverted Ordering}
\label{sub:3nuIO}

In the case of Inverted Ordering (IO),
the expressions of the neutrino masses in terms of the lightest neutrino mass
$m_{\text{min}}$ are:
\begin{align}
&
m_{3} = m_{\text{min}}
,
\label{m3IO3nu}
\\
&
m_{1} = \sqrt{m_{\text{min}}^2 + \Delta{m}_{\text{ATM}}^2 - \Delta{m}_{\text{SOL}}^2/2}
,
\label{m1IO3nu}
\\
&
m_{2} = \sqrt{m_{\text{min}}^2 + \Delta{m}_{\text{ATM}}^2 + \Delta{m}_{\text{SOL}}^2/2}
.
\label{m2IO3nu}
\end{align}

Figure~\ref{3nuIOpartial}
shows the best-fit values
and the
$1\sigma$,
$2\sigma$ and
$3\sigma$ allowed intervals
of the three partial mass contributions to $|m_{\beta\beta}|$ in Eq.~(\ref{mbb3nu})
as functions of
the lightest mass $m_{3}$.
One can see that $\prt_{1}$ is always dominant,
because $\vartheta_{12}$ is smaller than $\pi/4$
and
$|U_{e1}| > |U_{e2}| > |U_{e3}|$.
Therefore,
in the case of Inverted Ordering
there cannot be a complete cancellation of the three mass contributions to
$|m_{\beta\beta}|$
(Fig.~\ref{ratIOmbbvsmin} shows that
$m_{\beta\beta}^{(-)}$
is always positive)
and we obtain
from Fig.~\ref{3nuIOmbbvsmin}
the lower bounds
\begin{equation}
|m_{\beta\beta}|
>
1.6
\, (1\sigma)
,
1.5
\, (2\sigma)
,
1.3
\, (3\sigma)
\times 10^{-2} \, \text{eV}
.
\label{mbbminIO3nu}
\end{equation}
In the case of an Inverted Hierarchy ($m_{3} \ll m_{1} < m_{2}$)
we also have the upper bounds
\begin{equation}
|m_{\beta\beta}|
<
4.8
\, (1\sigma)
,
4.9
\, (2\sigma)
,
4.9
\, (3\sigma)
\times 10^{-2} \, \text{eV}
.
\label{mbbmaxIO3nu}
\end{equation}
The next generations of
neutrinoless double-beta decay experiments
(see Refs.~\cite{GomezCadenas:2011it,Giuliani:2012zu,Schwingenheuer:2012zs,Cremonesi:2013vla,Artusa:2014wnl,Gomez-Cadenas:2015twa})
will try to explore the range of $|m_{\beta\beta}|$
between the limits in Eqs.~(\ref{mbbminIO3nu}) and (\ref{mbbmaxIO3nu}),
testing the Majorana nature of neutrinos
in the case of an Inverted Hierarchy.

Figures~\ref{3nuIOmbbvsmb} and \ref{3nuIOmbbvssum}
show the correlation between $|m_{\beta\beta}|$
and the measurable quantities
$m_{\beta}$ and $\Sigma$
in the Inverted Ordering.
Since in this case both
$m_{\beta}$ and $\Sigma$
have relatively large lower bounds
($4.6 \times 10^{-2} \, \text{eV}$
and
$9.4 \times 10^{-2} \, \text{eV}$,
respectively, at $3\sigma$)
there is a concrete possibility that near-future experiments
will determine an allowed region in these plots
if in nature there are only three neutrinos with Inverted Ordering.

\section{3+1 Mixing}
\label{sec:3p1}

In this section we consider the case of 3+1 mixing
in which there is a new massive neutrino $\nu_{4}$ at the eV scale which is mainly sterile.
As explained in Section~\ref{sec:intro},
3+1 mixing
is motivated
\cite{Kopp:2013vaa,Giunti:2013aea}
by the explanation of the
reactor, Gallium and LSND
anomalies,
which requires the existence of a new squared-mass difference
$\Delta{m}^2_{\text{SBL}} \sim 1 \, \text{eV}^2$.
In this case,
the effective Majorana mass in neutrinoless double-beta decay
is given by
\begin{equation}
|m_{\beta\beta}|
=
\left| \prt_{1} + \prt_{2} e^{i\alpha_2} + \prt_{3} e^{i\alpha_3}  + \prt_{4} e^{i\alpha_4} \right|
,
\label{mbb3p1}
\end{equation}
with the partial mass contributions given by Eq.~(\ref{mkp}).
The contribution of $\nu_{4}$
enters with a totally unknown new phase
$\alpha_4$
that must be varied from 0 to $2\pi$
as
$\alpha_2$ and $\alpha_3$
in order to calculate the predictions of the value of
$|m_{\beta\beta}|$.

The absolute  values of the relevant first row of the $4\times4$
mixing matrix $U$
is given by the simple extension of the standard parameterization:
\begin{align}
&
|U_{e1}| = \cos\vartheta_{14} \cos\vartheta_{13} \cos\vartheta_{12},
\label{3p1Ue1}
\\
&
|U_{e2}| = \cos\vartheta_{14} \cos\vartheta_{13} \sin\vartheta_{12},
\label{3p1Ue2}
\\
&
|U_{e3}| = \cos\vartheta_{14} \sin\vartheta_{13},
\label{3p1Ue3}
\\
&
|U_{e4}| = \sin\vartheta_{14}.
\label{3p1Ue4}
\end{align}

Since in the case of 3+1 neutrino mixing,
as well as in any extension of the standard $3\nu$ mixing,
the ordering of the three standard massive neutrinos is not known,
in the following two subsections
we consider separately the two cases
of Normal and Inverted Ordering of
$\nu_{1}$,
$\nu_{2}$,
$\nu_{3}$.
The values of their masses
as functions of the lightest mass
$m_{\text{min}}$
are given by
Eqs.~(\ref{m1NO3nu})--(\ref{m3NO3nu}) in the Normal Ordering
and by
Eqs.~(\ref{m3IO3nu})--(\ref{m2IO3nu}) in the Inverted Ordering.
However,
in both cases we have
\begin{equation}
m_{4}
\simeq
\sqrt{m_{\text{min}}^2 + \Delta{m}^2_{\text{SBL}}}
,
\label{m4}
\end{equation}
neglecting
the contributions of
$\Delta{m}^2_{\text{SOL}}$
and
$\Delta{m}^2_{\text{ATM}}$,
which are much smaller than
$\Delta{m}^2_{\text{SBL}}$.
We calculated the confidence intervals using the $\chi^2$ function
\begin{equation}
\chi^2_{3+1}
=
\chi^2_{3\nu}
+
\chi^2(\Delta{m}^2_{\text{SBL}}, \sin^2\vartheta_{14})
,
\label{chi3p1}
\end{equation}
with
$\chi^2_{3\nu}$ defined in Eq.~(\ref{chi3nu})
and
$\chi^2(\Delta{m}^2_{\text{SBL}}, \sin^2\vartheta_{14})$
obtained from an update \cite{Giunti-NeuTel2015,Gariazzo:2015rra}
of the global fit of short-baseline neutrino oscillation
data presented in Ref.~\cite{Giunti:2013aea}.

After Eq.~(\ref{chi3nu}) we noted that in the case of $3\nu$ mixing
our statistical method for the calculation of the uncertainty of
$|m_{\beta\beta}|$
and the usual method based on the
propagation of errors lead to similar results,
because the $\chi^2$'s of the relevant $3\nu$ mixing parameters
are very well approximated by quadratic functions.
On the other hand,
the usual propagation of errors is inaccurate in the case of 3+1 mixing,
because the marginal $\chi^2$'s of
$\Delta{m}^2_{\text{SBL}}$
and
$\sin^2\vartheta_{14}$
are not quadratic.
Moreover,
there are significant correlations between
$\Delta{m}^2_{\text{SBL}}$
and
$\sin^2\vartheta_{14}$
(see Fig.~3 of Ref.~\cite{Giunti:2013aea})
which are taken into account in
$\chi^2(\Delta{m}^2_{\text{SBL}}, \sin^2\vartheta_{14})$.

\begin{figure}[t!]
\centering
\includegraphics*[width=\linewidth]{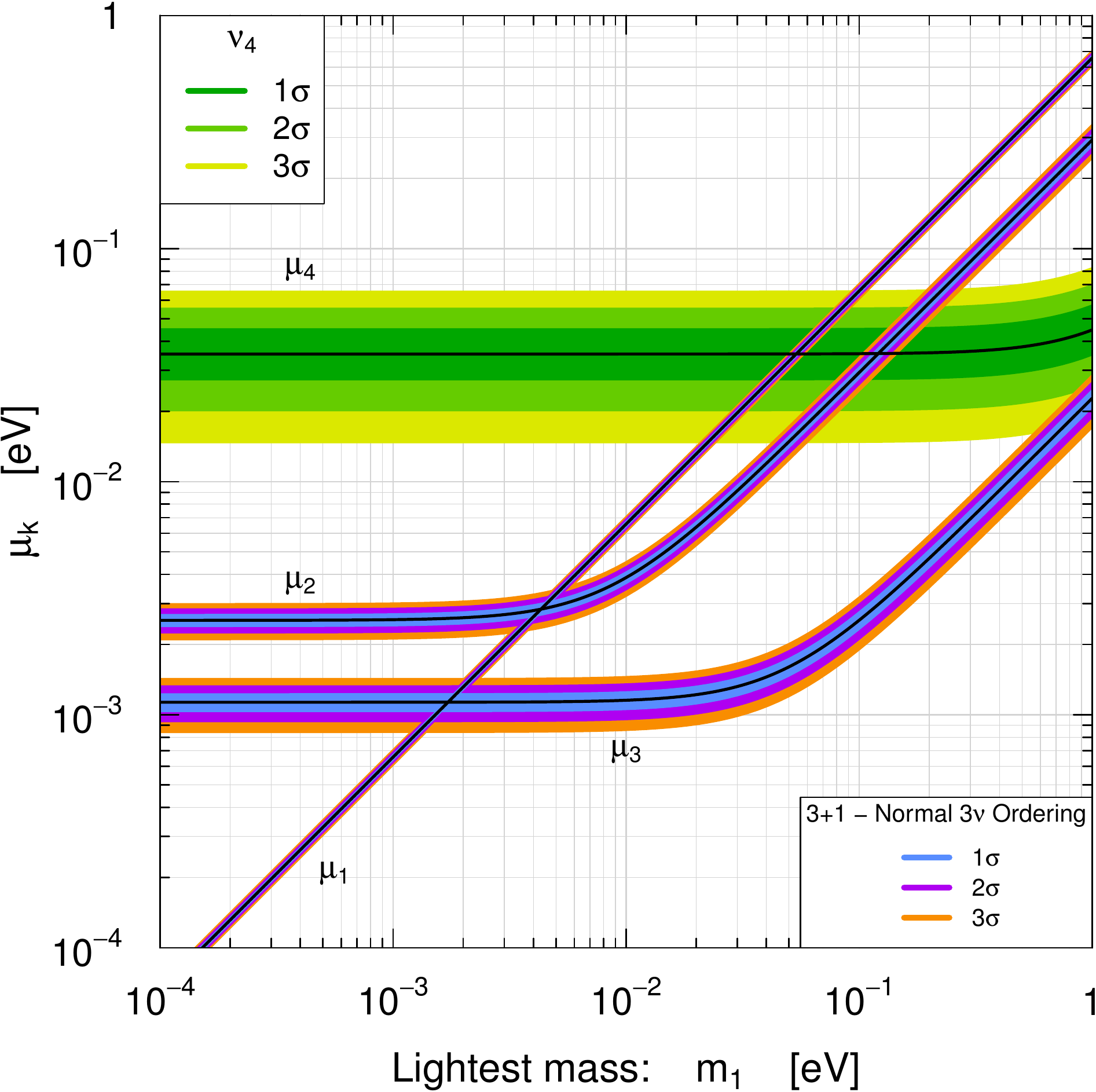}
\caption{Best-fit values (b.f.)
and
$1\sigma$,
$2\sigma$ and
$3\sigma$ allowed intervals
of the four partial mass contributions to $|m_{\beta\beta}|$ in Eq.~(\ref{mbb3p1})
as functions of
the lightest mass $m_{1}$
in the case of 3+1 mixing with Normal Ordering
of the three lightest neutrinos.}
\label{3p1NOpartial}
\end{figure}

\begin{figure}[t!]
\centering
\includegraphics*[width=\linewidth]{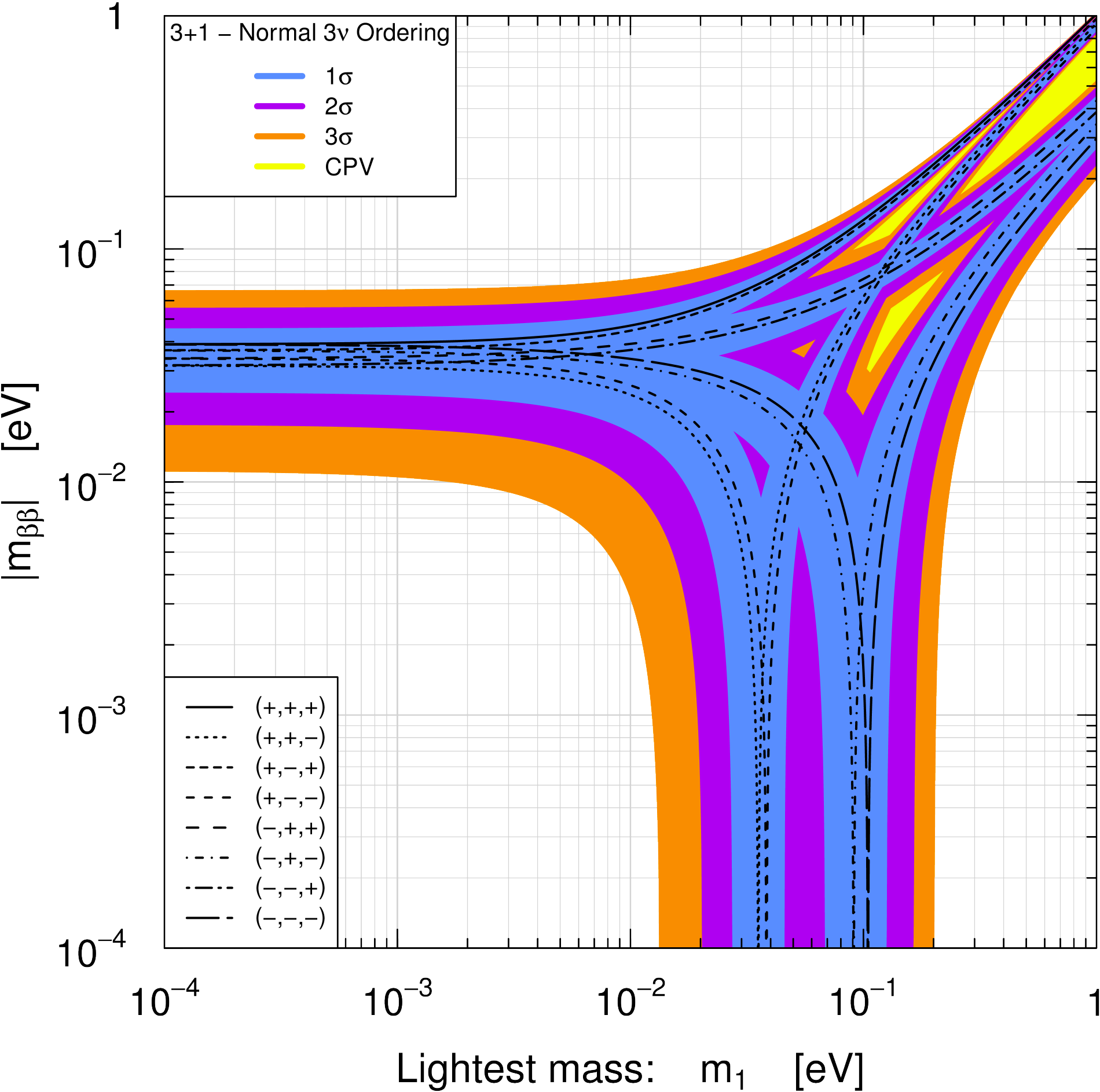}
\caption{
Value of the effective Majorana mass $|m_{\beta\beta}|$
as a function of the lightest neutrino mass $m_{1}$
in the case of 3+1 mixing with Normal Ordering
of the three lightest neutrinos.
The signs in the legend indicate the signs of
$e^{i\alpha_{2}}, e^{i\alpha_{3}} , e^{i\alpha_{4}} = \pm1$
for the four possible cases in which CP is conserved.
The intermediate yellow region is allowed only in the case of CP violation.
}
\label{3p1NOmbbvsmin}
\end{figure}

\begin{figure}[t!]
\centering
\includegraphics*[width=\linewidth]{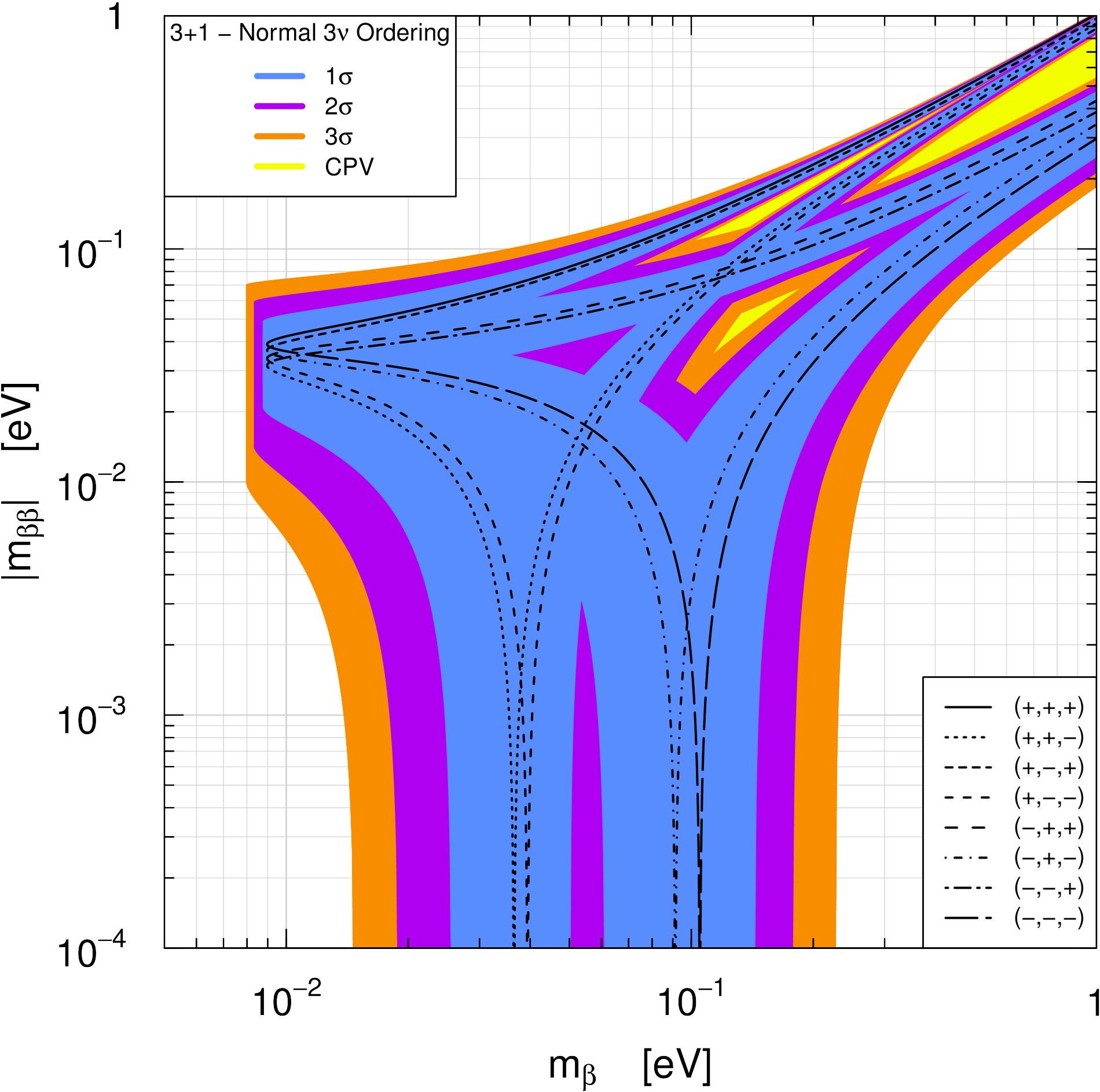}
\caption{
Value of the effective Majorana mass $|m_{\beta\beta}|$
as a function of effective electron neutrino mass $m_{\beta}$ in Eq.~(\ref{mb})
in the case of 3+1 mixing with Normal Ordering
of the three lightest neutrinos.
The legend is explained in the caption of Fig.~\ref{3p1NOmbbvsmin}.
}
\label{3p1NOmbbvsmb}
\end{figure}

\begin{figure}[t!]
\centering
\includegraphics*[width=\linewidth]{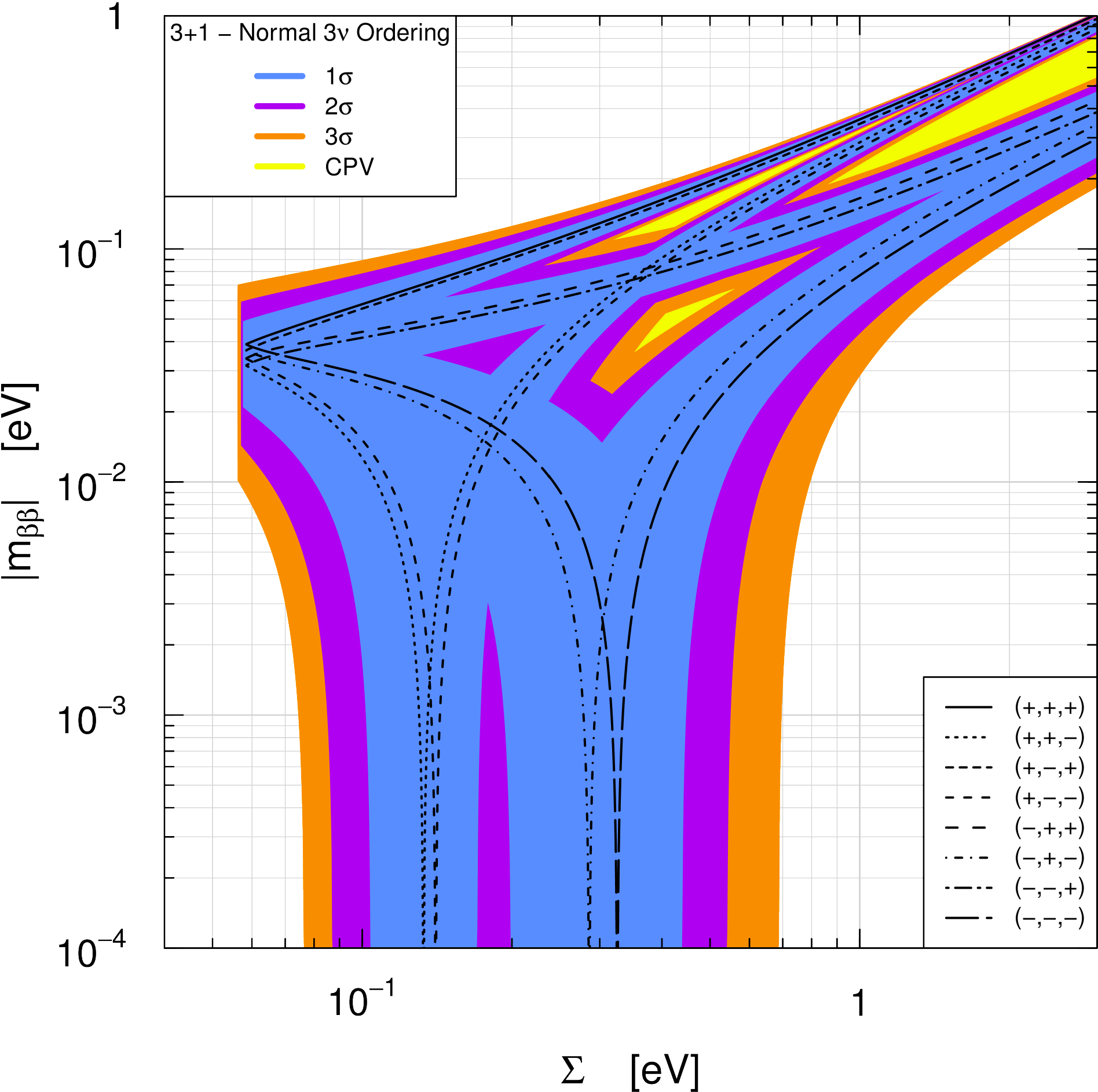}
\caption{
Value of the effective Majorana mass $|m_{\beta\beta}|$
as a function of sum of the three lightest neutrino masses $\Sigma$ in Eq.~(\ref{sum})
in the case of 3+1 mixing with Normal Ordering
of the three lightest neutrinos.
The legend is explained in the caption of Fig.~\ref{3p1NOmbbvsmin}.
}
\label{3p1NOmbbvssum}
\end{figure}

\begin{figure}[t!]
\centering
\includegraphics*[width=\linewidth]{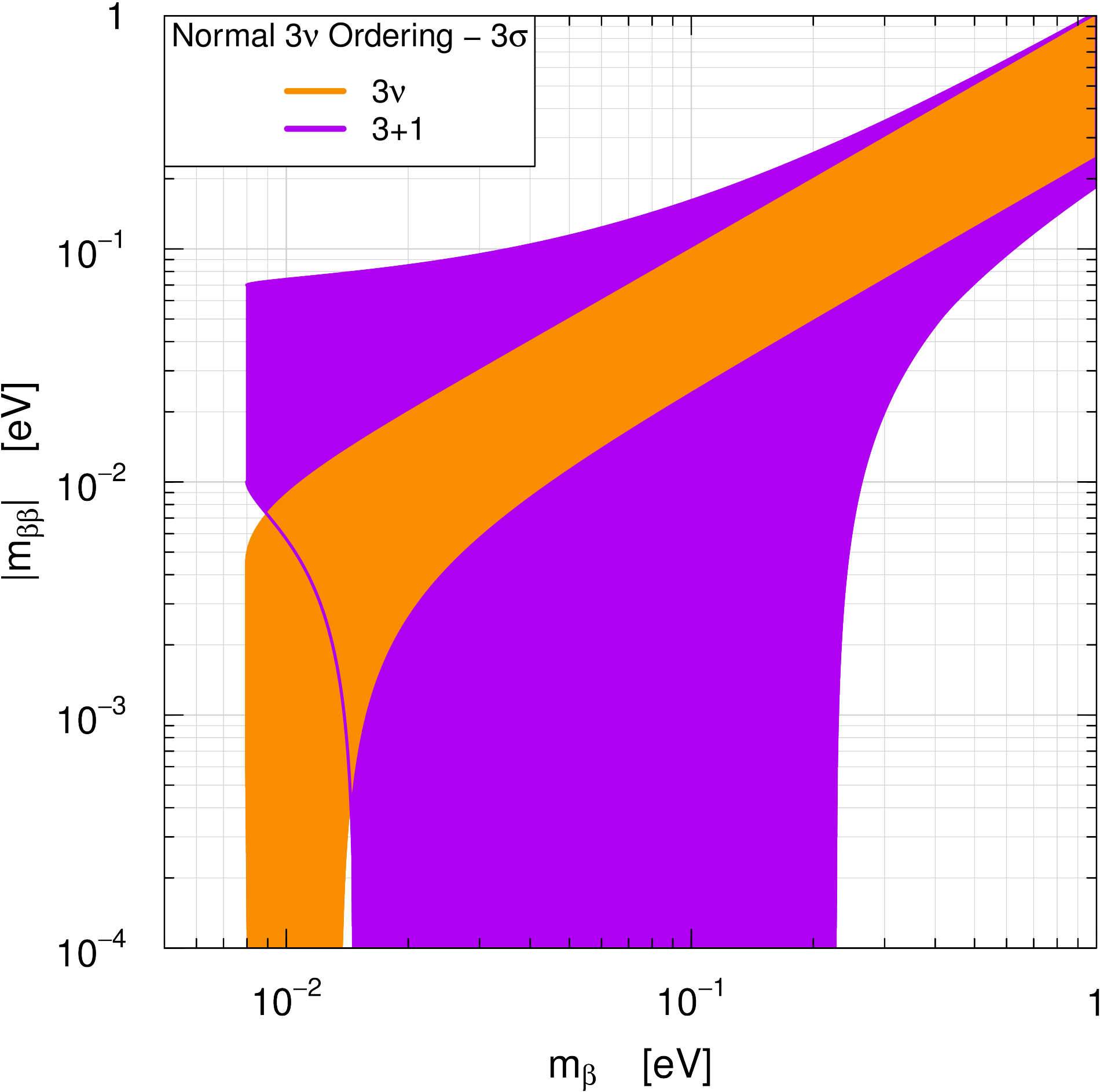}
\caption{Comparison of the $3\sigma$ allowed regions in the
$m_{\beta}$--$|m_{\beta\beta}|$ plane
in the cases of $3\nu$ and 3+1 mixing with Normal Ordering
of the three lightest neutrinos.}
\label{compNOmbbvsmb}
\end{figure}

\begin{figure}[t!]
\centering
\includegraphics*[width=\linewidth]{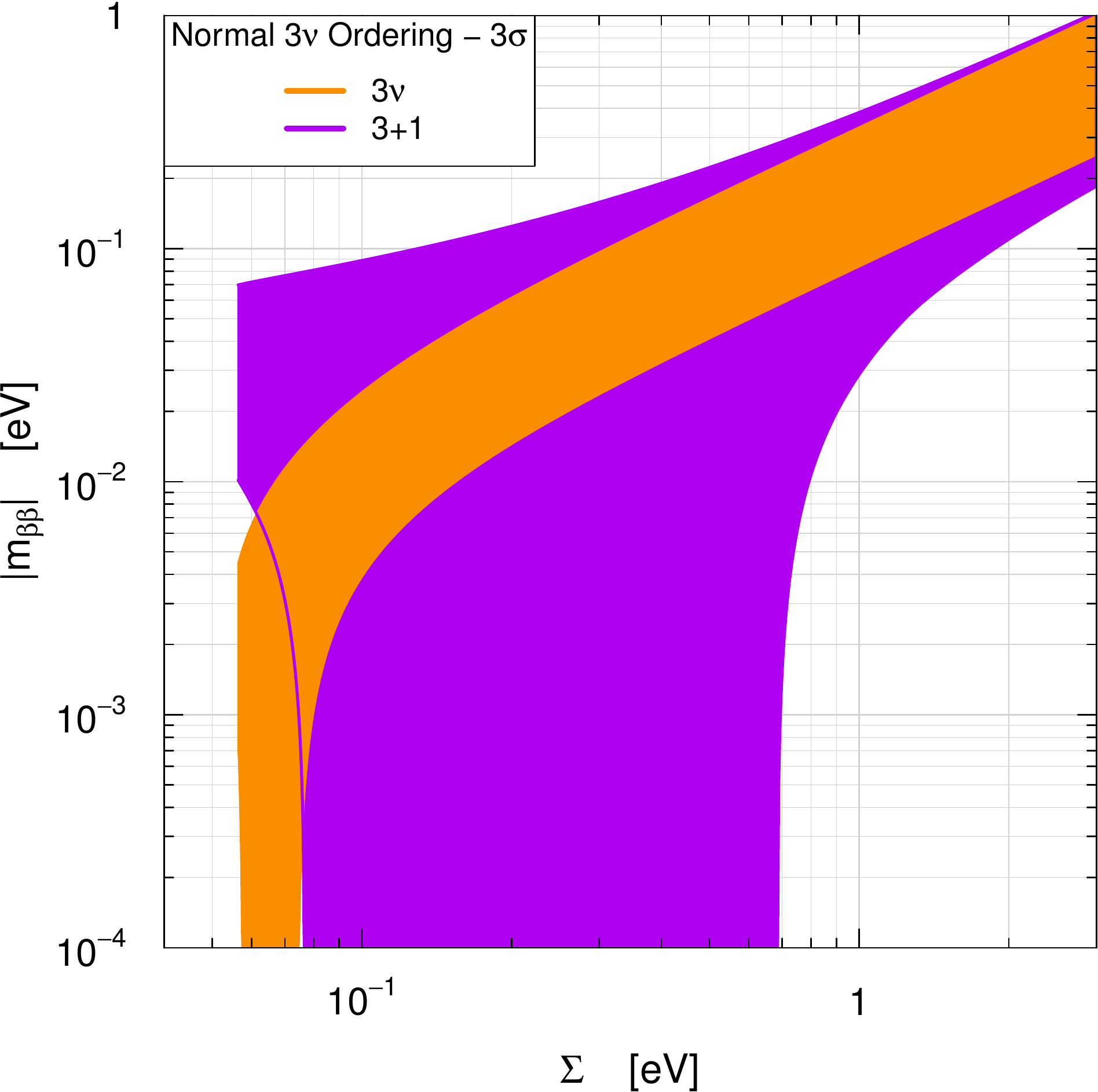}
\caption{Comparison of the $3\sigma$ allowed regions in the
$\Sigma$--$|m_{\beta\beta}|$ plane
in the cases of $3\nu$ and 3+1 mixing with Normal Ordering
of the three lightest neutrinos.}
\label{compNOmbbvssum}
\end{figure}

\begin{table}[b!]
\caption{Ranges of
$m_{1}$,
$m_{\beta}$ and
$\Sigma$
for which there can be a complete cancellation of the
four partial mass contributions to $|m_{\beta\beta}|$
for the best-fit values (b.f.) of the oscillation parameters
and at
$1\sigma$,
$2\sigma$ and
$3\sigma$
in the case of 3+1 mixing with Normal Ordering
of the three lightest neutrinos.
}
\centering
\renewcommand{\arraystretch}{1.2}
\begin{tabular}{ccccc}
\hline
\hline
&
b.f.
&
$1\sigma$
&
$2\sigma$
&
$3\sigma$
\\
\hline
$m_{1}\,[10^{-2}\,\text{eV}]$
&
$3.5 - 10.5$
&
$2.8 - 12.5$
&
$2.0 - 16.3$
&
$1.3 - 20.0$
\\
$m_{\beta}\,[10^{-2}\,\text{eV}]$
&
$3.6 - 10.5$
&
$2.6 - 14.3$
&
$1.9 - 17.8$
&
$1.5 - 22.7$
\\
$\Sigma\,[10^{-2}\,\text{eV}]$
&
$13.3 - 32.6$
&
$10.4 - 43.8$
&
$8.7 - 53.9$
&
$7.6 - 68.5$
\\
\hline
\hline
\end{tabular}
\label{tab:3p1-nor}
\end{table}

\subsection{Normal Ordering}
\label{sub:3p1NO}

Figure~\ref{3p1NOpartial}
shows a comparison of the best-fit value
and the
$1\sigma$,
$2\sigma$ and
$3\sigma$ allowed intervals
of the partial contribution
$\prt_{4}$
as a function of
the lightest mass $m_{1}$
with those of
$\prt_{1}$,
$\prt_{2}$,
$\prt_{3}$,
which are slightly different from those in Fig.~\ref{3nuNOpartial}
because of the contribution of $\vartheta_{14}$ in Eqs.~(\ref{3p1Ue1})--(\ref{3p1Ue3}).
One can see that in the 3+1 case it is not possible to get a total cancellation of
$|m_{\beta\beta}|$
in the interval
$ m_{1} \approx (2-7) \times 10^{-3} \, \text{eV} $
as in the case of $3\nu$ mixing
(see Tab.~\ref{tab:3nu-nor}),
because in this interval of $m_{1}$
the contribution of
$\prt_{4}$
is dominant.
However,
there is a range of higher values of $m_{1}$
between about
0.02 and 0.2 eV
in which $\prt_{4}$ and $\prt_{1}$
have similar values,
leading to a possible total cancellation.
For $ m_{1} \gtrsim 0.2 \, \text{eV} $
a total cancellation is again not possible
because the contribution of $\prt_{1}$ is dominant.
This behavior is confirmed by Fig.~\ref{ratNOmbbvsmin},
where one can see that the value of
$m_{\beta\beta}^{(-)}$ (see Eq.~(\ref{mbbpm}))
corresponding to the best-fit values of the
partial mass contributions
is negative for
$
0.035
\lesssim
m_{1}
\lesssim
0.1
\, \text{eV}
$.

Figure~\ref{3p1NOmbbvsmin}
shows the allowed values of
$|m_{\beta\beta}|$
as a function of $m_{1}$
at different confidence levels.
The corresponding intervals of
$m_{1}$ for which there can be a total cancellation of
$|m_{\beta\beta}|$
are given in Tab.~\ref{tab:3p1-nor}.

In Fig.~\ref{3p1NOmbbvsmin}
we have plotted separately the
allowed bands for the eight possible cases in which CP is conserved
($\alpha_{2}, \alpha_{3}, \alpha_{4} = 0, \pi$),
that are the extreme cases which determine the minimum and maximum values
of $|m_{\beta\beta}|$.
The areas between the CP-conserving allowed bands
correspond to values of
$|m_{\beta\beta}|$ which are allowed only in the case of CP violation.
Unfortunately,
these areas are visibly smaller than those in Fig.~\ref{3nuNOmbbvsmin}
in the case of $3\nu$ mixing.
This is due to the relatively large uncertainty of
$\prt_{4}$,
which can be seen clearly in Fig.~\ref{3p1NOpartial}.
This uncertainty broadens the allowed bands
corresponding to the CP-conserving cases,
leaving little intermediate space.
In any case,
even if the uncertainty of
$\prt_{4}$
will be reduced in the future,
there cannot be a region which is allowed only in the case of CP-violation for
$ m_{1} \lesssim 10^{-2} \, \text{eV} $,
where
$\prt_{4}$ is dominant
and the CP-violating phases are irrelevant.
In fact, all the best-fit CP-conserving curves have approximately the same value
for
$ m_{1} \lesssim 10^{-2} \, \text{eV} $.

As in the case of $3\nu$ mixing,
the plot in Fig.~\ref{3p1NOmbbvsmin}
of
$|m_{\beta\beta}|$ as a function of the lightest mass $m_{1}$
is useful because it gives a clear view of the different possibilities for the value
of $|m_{\beta\beta}|$,
but in practice it will be very difficult to determine experimentally
an allowed region in this plot because of the difficulty of
measuring the value of the lightest mass.
Therefore,
we calculated also the allowed regions in the
$m_{\beta}$--$|m_{\beta\beta}|$
and
$\Sigma$--$|m_{\beta\beta}|$
planes shown in
Figs.~\ref{3p1NOmbbvsmb} and \ref{3p1NOmbbvssum},
with the quantities
$m_{\beta}$
and
$\Sigma$
defined in Eqs.~(\ref{mb}) and (\ref{sum})
as in the case of $3\nu$ mixing
in terms of the three standard neutrino masses only.
The reason of this choice is that
$m_{\beta}$
and
$\Sigma$
are measurable quantities also in the 3+1 scheme.
Indeed,
considering $\beta$ decay,
$m_{\beta}$
quantifies approximately the deviation of the end-point
of the electron spectrum due to neutrino masses smaller than
the experimental energy resolution
\cite{Shrock:1980vy,McKellar:1980cn,Kobzarev:1980nk,Vissani:2000ci,Farzan:2002zq},
whereas
the effect of the larger mass $m_{4}$
is a kink of the Kurie function
(see Ref.~\cite{Giunti:2007ry}).
In cosmology,
the effects of the larger mass $m_{4}$
can be disentangled from those of the smaller masses,
because $\nu_{4}$ becomes non-relativistic
shortly after matter-radiation equality,
much earlier than
$\nu_{1}$,
$\nu_{2}$,
$\nu_{3}$.
Moreover,
it is possible that the contribution of
$m_{4}$
to the energy density of the Universe
is suppressed,
for example by a large lepton asymmetry
\cite{Chu:2006ua,Hannestad:2012ky,Mirizzi:2012we,Saviano:2013ktj,Hannestad:2013pha},
or an enhanced background potential due to new interactions in the sterile sector
\cite{Hannestad:2013ana,Dasgupta:2013zpn,Bringmann:2013vra,Ko:2014bka,Archidiacono:2014nda,Saviano:2014esa,Mirizzi:2014ama},
or a larger cosmic expansion rate at the time of sterile neutrino production
\cite{Rehagen:2014vna},
or MeV dark matter annihilation
\cite{Ho:2012br}.

From Figs.~\ref{3p1NOmbbvsmb} and \ref{3p1NOmbbvssum}
one can see that the intervals of
$m_{\beta}$ and $\Sigma$
for which there can be a complete cancellation of the
three partial mass contributions to $|m_{\beta\beta}|$
(given in Tab.~\ref{tab:3p1-nor})
are much larger than those in
Figs.~\ref{3nuNOmbbvsmb} and \ref{3nuNOmbbvssum}
for the standard $3\nu$ mixing case,
and
$|m_{\beta\beta}| \gtrsim 0.01 \,\text{eV}$
for any value of the unknown phases
$\alpha_2$,
$\alpha_3$,
$\alpha_4$
only for the relatively large values
$m_{\beta} \gtrsim 0.25 \, \text{eV}$
and
$\Sigma \gtrsim 0.8 \, \text{eV}$.

It is useful to compare the allowed regions
$m_{\beta}$--$|m_{\beta\beta}|$
and
$\Sigma$--$|m_{\beta\beta}|$
planes obtained
in the cases of $3\nu$ and 3+1 mixing with Normal Ordering
of the three lightest neutrinos.
Figures~\ref{compNOmbbvsmb} and \ref{compNOmbbvssum}
show this comparison for the $3\sigma$ allowed regions.
One can see that,
if the Normal Ordering will be established by oscillation experiments
(see Refs.~\cite{Bellini:2013wra,Wang:2015rma}),
with measurements of
$m_{\beta}$ and $|m_{\beta\beta}|$
and/or
$\Sigma$ and $|m_{\beta\beta}|$
it may be possible to distinguish $3\nu$ mixing and 3+1 mixing
if the measured values
select a region which is allowed only in one of the two cases.
It is interesting that there are two regions allowed only to 3+1 mixing:
one with $|m_{\beta\beta}|$ smaller than that in the case of $3\nu$ mixing
and
one with $|m_{\beta\beta}|$ larger than that in the case of $3\nu$ mixing.
At least a part of the second region is accessible to the next generation of
neutrinoless double-beta decay experiments
(see Refs.~\cite{GomezCadenas:2011it,Giuliani:2012zu,Schwingenheuer:2012zs,Cremonesi:2013vla,Artusa:2014wnl,Gomez-Cadenas:2015twa}).
The only $\beta$-decay experiment under preparation
with the aim of exploring the sub-eV region
of $m_{\beta}$
is KATRIN \cite{fortheKATRIN:2013saa},
which will have a sensitivity of about 0.2 eV
that is not sufficient to explore the upper part of the
region in Fig.~\ref{compNOmbbvsmb}
allowed only in the case of 3+1 mixing.
On the other hand,
cosmological observation may
be able to measure the sum of the three light neutrino masses down to the lower limit
of about
$5.6 \times 10^{-2} \, \text{eV}$
\cite{Audren:2012vy}.

\begin{figure}[t!]
\centering
\includegraphics*[width=\linewidth]{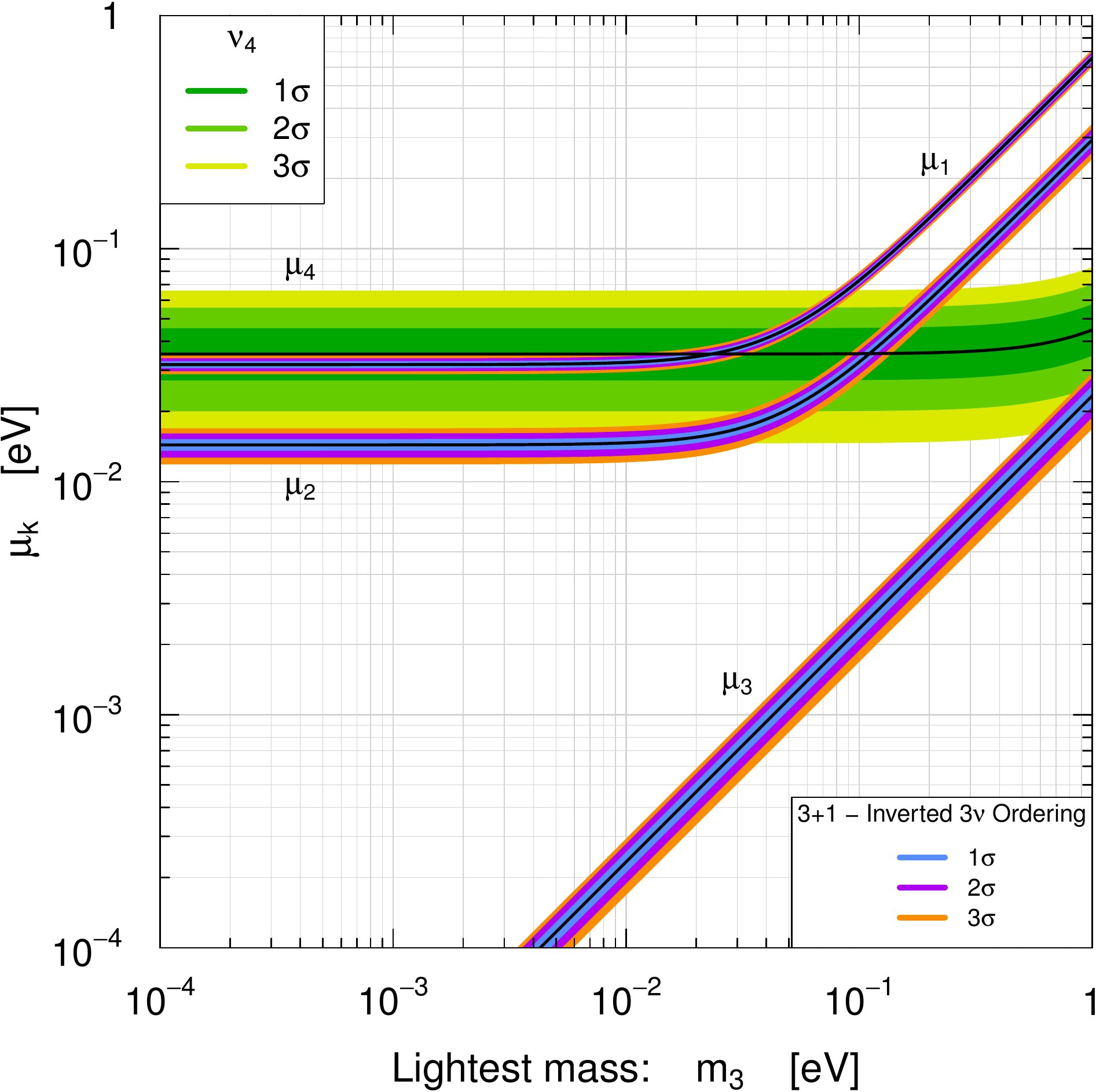}
\caption{
Best-fit values (b.f.)
and
$1\sigma$,
$2\sigma$ and
$3\sigma$ allowed intervals
of the four partial mass contributions to $|m_{\beta\beta}|$ in Eq.~(\ref{mbb3p1})
as functions of
the lightest mass $m_{3}$
in the case of 3+1 mixing with Inverted Ordering
of the three lightest neutrinos.
}
\label{3p1IOpartial}
\end{figure}

\begin{figure}[t!]
\centering
\includegraphics*[width=\linewidth]{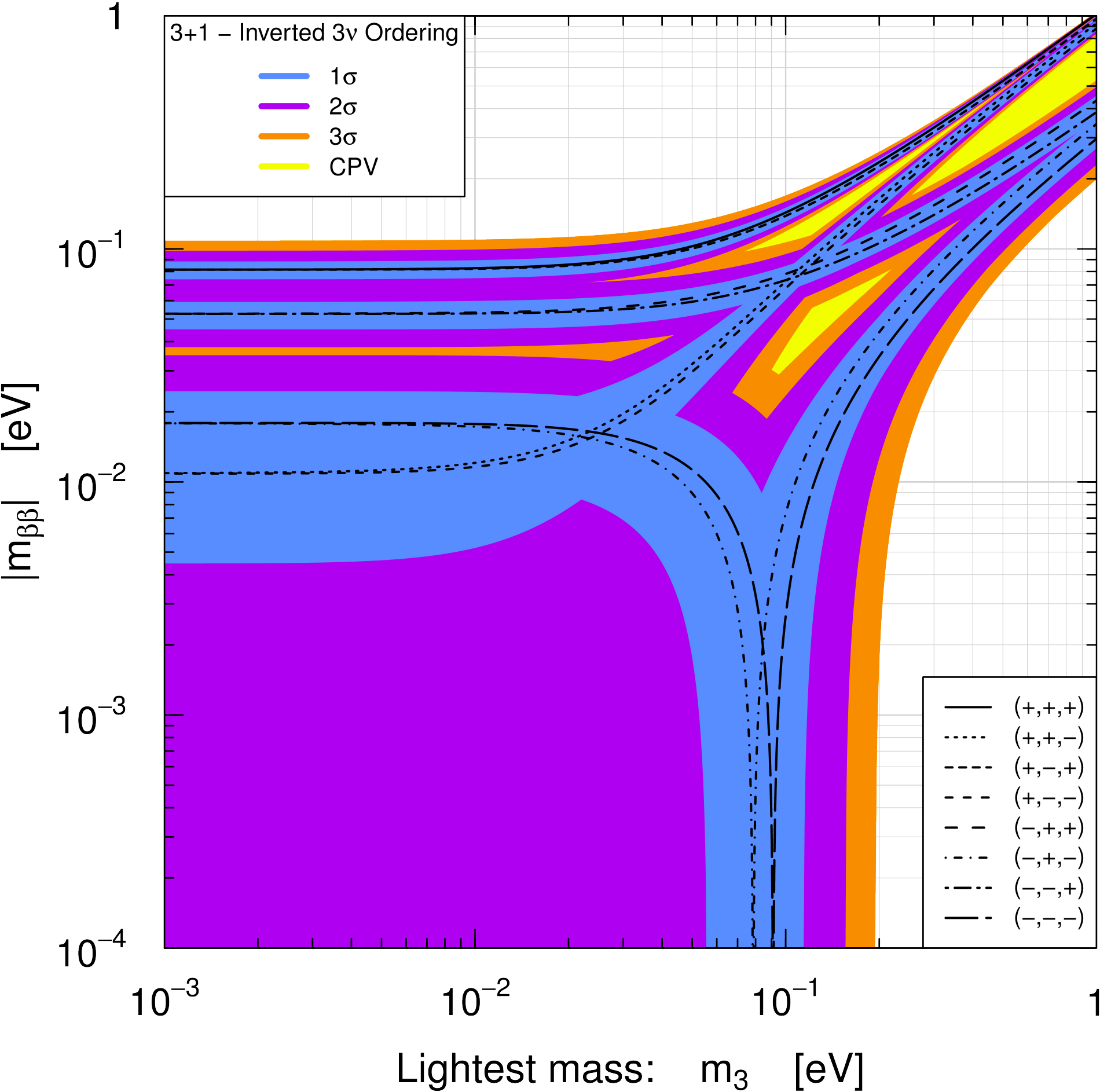}
\caption{
Value of the effective Majorana mass $|m_{\beta\beta}|$
as a function of the lightest neutrino mass $m_{3}$
in the case of 3+1 mixing with Inverted Ordering
of the three lightest neutrinos.
The legend is explained in the caption of Fig.~\ref{3p1NOmbbvsmin}.
}
\label{3p1IOmbbvsmin}
\end{figure}

\begin{figure}[t!]
\centering
\includegraphics*[width=\linewidth]{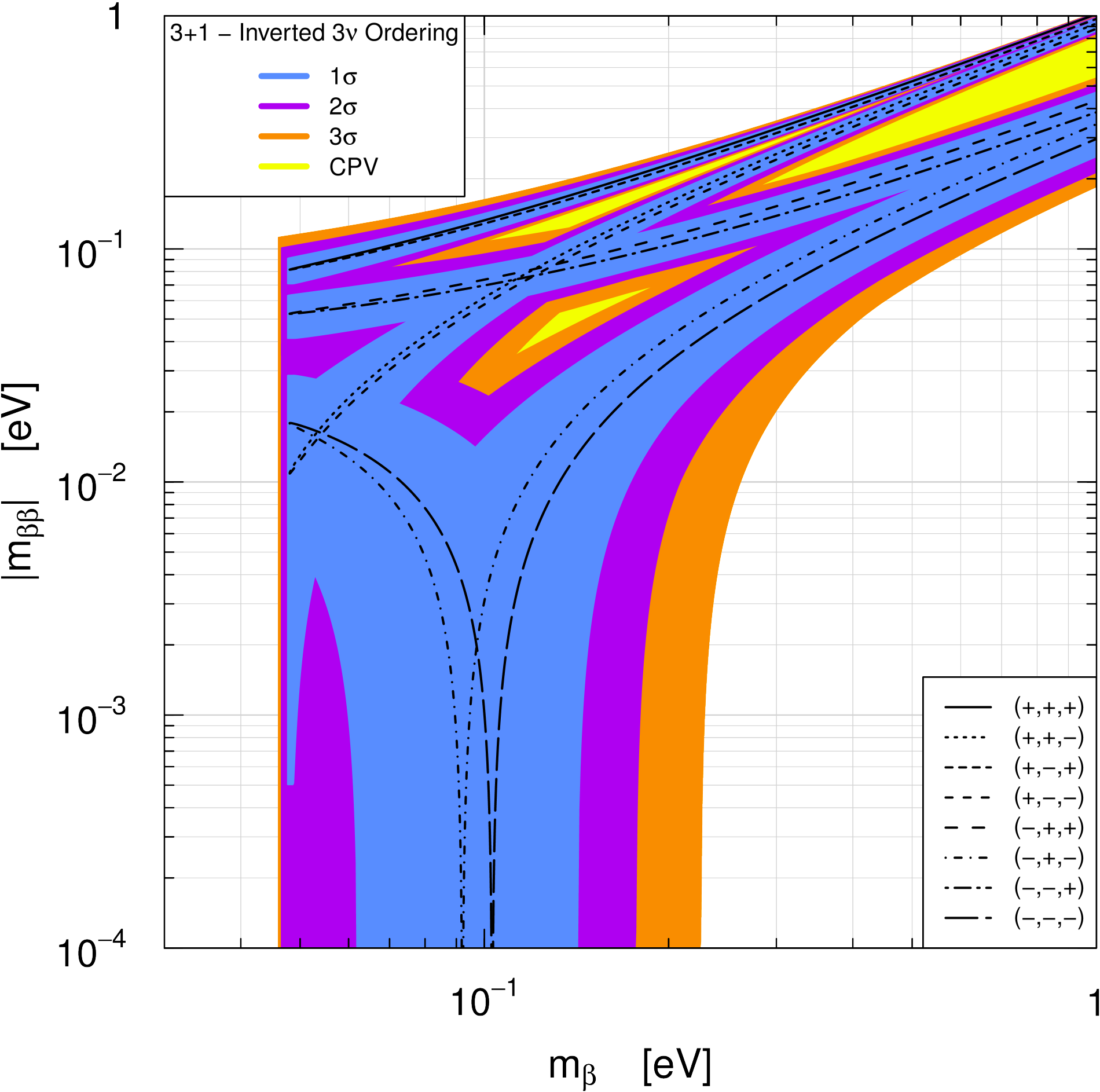}
\caption{
Value of the effective Majorana mass $|m_{\beta\beta}|$
as a function of sum of the three lightest neutrino masses $\Sigma$ in Eq.~(\ref{sum})
in the case of 3+1 mixing with Inverted Ordering
of the three lightest neutrinos.
The legend is explained in the caption of Fig.~\ref{3p1NOmbbvsmin}.
}
\label{3p1IOmbbvsmb}
\end{figure}

\begin{figure}[t!]
\centering
\includegraphics*[width=\linewidth]{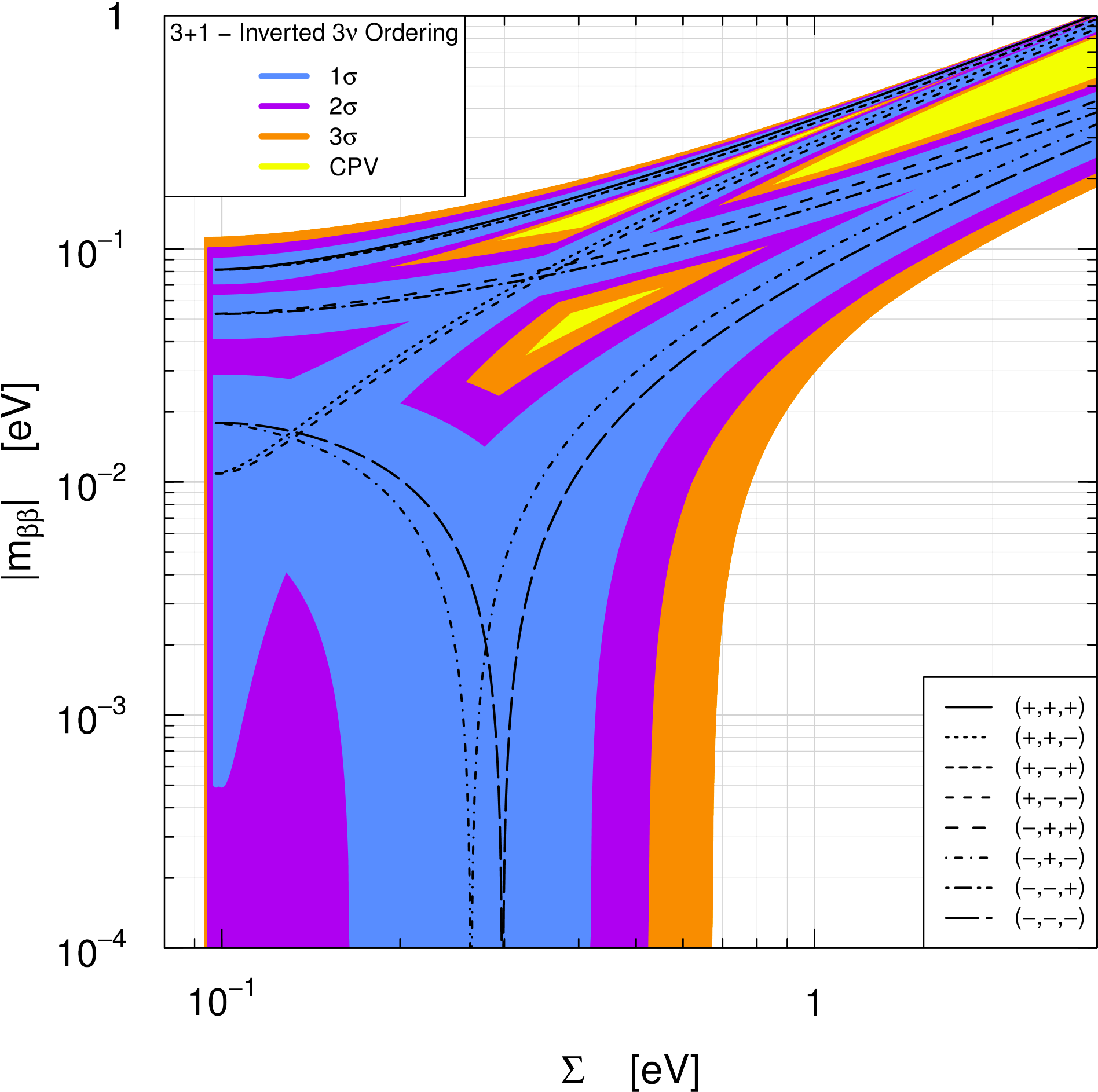}
\caption{Value of the effective Majorana mass $|m_{\beta\beta}|$
as a function of effective electron neutrino mass $m_{\beta}$ in Eq.~(\ref{mb})
in the case of 3+1 mixing with Inverted Ordering
of the three lightest neutrinos.
The legend is explained in the caption of Fig.~\ref{3p1NOmbbvsmin}.
}
\label{3p1IOmbbvssum}
\end{figure}

\begin{table}[b!]
\caption{Ranges of
$m_{3}$,
$m_{\beta}$ and
$\Sigma$
for which there can be a complete cancellation of the
four partial mass contributions to $|m_{\beta\beta}|$
for the best-fit values (b.f.) of the oscillation parameters
and at
$1\sigma$,
$2\sigma$ and
$3\sigma$
in the case of 3+1 mixing with Inverted Ordering
of the three lightest neutrinos.
}
\centering
\renewcommand{\arraystretch}{1.2}
\begin{tabular}{ccccc}
\hline
\hline
&
b.f.
&
$1\sigma$
&
$2\sigma$
&
$3\sigma$
\\
\hline
$m_{3}\,[10^{-2}\,\text{eV}]$
&
$< 9.1$
&
$< 11.4$
&
$< 15.5$
&
$< 19.3$
\\
$m_{\beta}\,[10^{-2}\,\text{eV}]$
&
$4.8 - 10.3$
&
$4.8 - 14.2$
&
$4.7 - 17.6$
&
$4.6 - 22.5$
\\
$\Sigma\,[10^{-2}\,\text{eV}]$
&
$9.8 - 29.9$
&
$9.7 - 41.8$
&
$9.5 - 52.3$
&
$9.4 - 67.1$
\\
\hline
\hline
\end{tabular}
\label{tab:3p1-inv}
\end{table}

\begin{figure}[t!]
\centering
\includegraphics*[width=\linewidth]{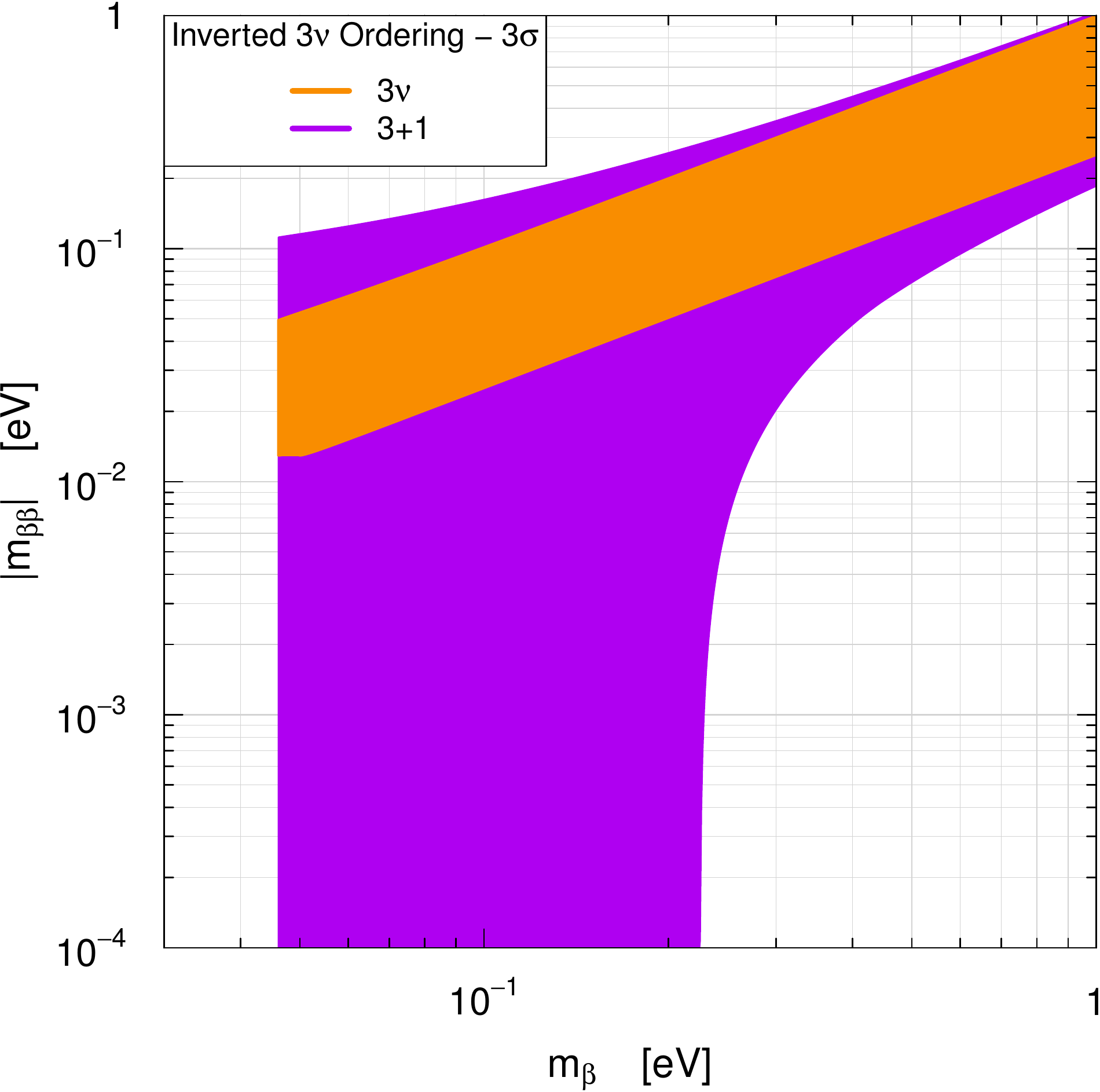}
\caption{Comparison of the $3\sigma$ allowed regions in the
$m_{\beta}$--$|m_{\beta\beta}|$ plane
in the cases of $3\nu$ and 3+1 mixing with Inverted Ordering
of the three lightest neutrinos.}
\label{compIOmbbvsmb}
\end{figure}

\begin{figure}[t!]
\centering
\includegraphics*[width=\linewidth]{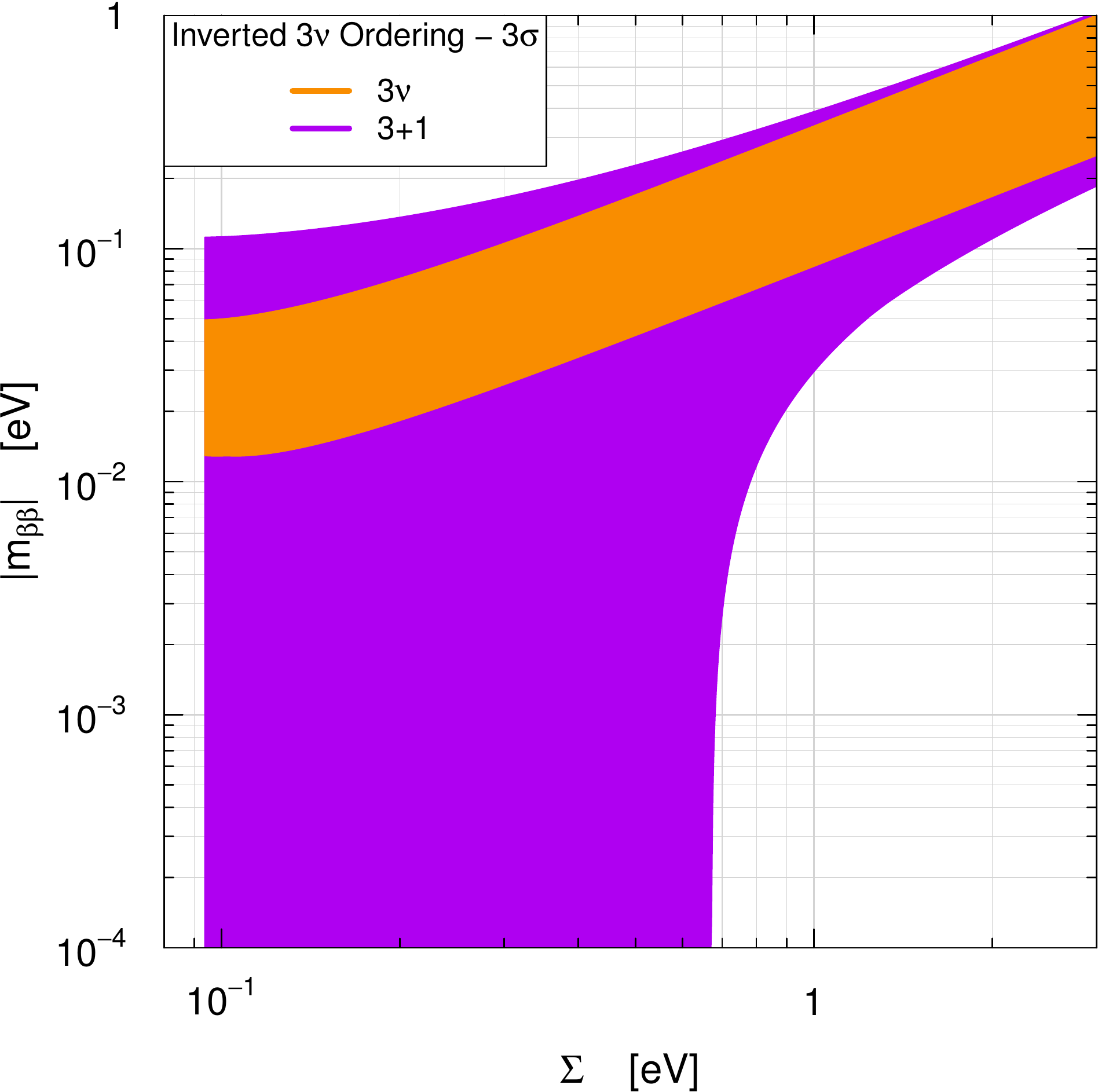}
\caption{Comparison of the $3\sigma$ allowed regions in the
$\Sigma$--$|m_{\beta\beta}|$ plane
in the cases of $3\nu$ and 3+1 mixing with Inverted Ordering
of the three lightest neutrinos.}
\label{compIOmbbvssum}
\end{figure}

\subsection{Inverted Ordering}
\label{sub:3p1IO}

The best-fit value
and the
$1\sigma$,
$2\sigma$ and
$3\sigma$ allowed intervals
of the partial mass contributions to $|m_{\beta\beta}|$
in the case of 3+1 mixing with Inverted Ordering
of the three lightest neutrinos
are shown in Fig.~\ref{3p1IOpartial}.
One can see that
there can be a total cancellation between the
partial contribution
$\prt_{4}$
and the dominant $\prt_{1}$ in the case of $3\nu$ mixing
(see Fig.~\ref{3nuIOpartial})
for $m_{3} \lesssim 0.1 \, \text{eV}$.
Indeed,
Fig.~\ref{ratIOmbbvsmin}
shows that the value of
$m_{\beta\beta}^{(-)}$ (see Eq.~(\ref{mbbpm}))
corresponding to the best-fit values of the
partial mass contributions
is negative for
$
m_{3}
\lesssim
0.09
\, \text{eV}
$.

Figure~\ref{3p1IOmbbvsmin}
depicts the allowed regions in
in the
$m_{3}$--$|m_{\beta\beta}|$ plane.
Comparing Fig.~\ref{3p1IOmbbvsmin}
with Fig.~\ref{3nuIOmbbvsmin}
one can see that
the predictions for $|m_{\beta\beta}|$
are completely different
in the $3\nu$ and 3+1 cases
if there is an Inverted Ordering
of the three lightest neutrinos,
in agreement with the discussions in
Refs.~\cite{Goswami:2005ng,Goswami:2007kv,Barry:2011wb,Li:2011ss,Rodejohann:2012xd,Giunti:2012tn,Girardi:2013zra,Pascoli:2013fiz,Meroni:2014tba,Abada:2014nwa}.
The ranges of values of $m_{3}$
for which there can be a complete cancellation of $|m_{\beta\beta}|$
are given in Tab.~\ref{tab:3p1-inv}.

Figures~\ref{3p1IOmbbvsmb} and \ref{3p1IOmbbvssum}
show the allowed regions in the
$m_{\beta}$--$|m_{\beta\beta}|$
and
$\Sigma$--$|m_{\beta\beta}|$
planes.
Figures~\ref{compIOmbbvsmb} and \ref{compIOmbbvssum}
show the comparison of the $3\sigma$ allowed regions
in the same planes
in the cases of $3\nu$ and 3+1 mixing with Inverted Ordering
of the three lightest neutrinos.
If the Inverted Ordering will be established by oscillation experiments
(see Refs.~\cite{Bellini:2013wra,Wang:2015rma}),
it will be possible to exclude $3\nu$ mixing in favor of 3+1
by restricting
$m_{\beta}$ and $|m_{\beta\beta}|$
or
$\Sigma$ and $|m_{\beta\beta}|$
in the corresponding large region at small $|m_{\beta\beta}|$
allowed only in the 3+1 case.

\section{Conclusions}
\label{sec:conclusions}

We have presented accurate calculations of the
effective Majorana mass $|m_{\beta\beta}|$
in neutrinoless double-$\beta$ decay
in the standard case of $3\nu$ mixing
and in the case of 3+1 neutrino mixing
indicated by the reactor, Gallium and LSND
anomalies
(see Refs.~\cite{Kopp:2013vaa,Giunti:2013aea}).
We have taken into account the uncertainties of the
standard $3\nu$ mixing parameters
obtained in the global fit
of solar, atmospheric and long-baseline reactor and accelerator
neutrino oscillation data presented in
Ref.~\cite{Capozzi:2013csa}
and the uncertainties on the additional mixing parameters
in the 3+1 case
obtained from an update \cite{Giunti-NeuTel2015,Gariazzo:2015rra}
of the global fit of short-baseline neutrino oscillation
data presented in Ref.~\cite{Giunti:2013aea}.

We have shown that the predictions for $|m_{\beta\beta}|$
in the cases of $3\nu$ and 3+1 mixing are quite different,
in agreement with the previous discussions in
Refs.~\cite{Goswami:2005ng,Goswami:2007kv,Barry:2011wb,Li:2011ss,Rodejohann:2012xd,Giunti:2012tn,Girardi:2013zra,Pascoli:2013fiz,Meroni:2014tba,Abada:2014nwa}.
Our paper improves these discussions
by taking into account the uncertainties of all the mixing parameters
and presenting all the results at
$1\sigma$,
$2\sigma$ and
$3\sigma$.

We have presented accurate comparisons of the
allowed regions in the planes
$m_{\beta}$--$|m_{\beta\beta}|$
and
$\Sigma$--$|m_{\beta\beta}|$
of measurable quantities,
taking into account the two possibilities of Normal and Inverted Ordering
of the three light lightest neutrinos.
We have shown that future measurements of these quantities
may distinguish the $3\nu$ and 3+1 cases
if the mass ordering is determined by oscillation experiments
(see Refs.~\cite{Bellini:2013wra,Wang:2015rma}).

We have also introduced
in Section~\ref{sub:3nuNO}
a relatively simple method
to determine the minimum value of
$|m_{\beta\beta}|$
in the general case of $N$-neutrino mixing.

\begin{acknowledgments}
E. Z. thanks the support of funding grants 2013/02518-7 and 2014/23980-3, S\~ao Paulo Research Foundation (FAPESP).
The work of C. Giunti is partially supported by the research grant {\sl Theoretical Astroparticle Physics} number 2012CPPYP7 under the program PRIN 2012 funded by the Ministero dell'Istruzione, Universit\`a e della Ricerca (MIUR).
\end{acknowledgments}


\end{document}